\documentclass[nofootinbib,reprint,amsmath,amssymb,prd,aps,superscriptaddress]{revtex4-2}

\usepackage{graphicx} 
\usepackage{dcolumn} 
\usepackage{hyperref} 
\usepackage{bm} 

\usepackage{physics}
\usepackage{xcolor}
\usepackage{subfigure}
\usepackage{hyperref}
\usepackage{ulem}
\usepackage{orcidlink}
\newcommand{\mev}{\text{\,MeV}}            
\newcommand{\msol}{\text{\,M$_\odot$}}     
\newcommand{\rhoo}{\text{\,$\rho_0$}}      
\newcommand{\ecc}{\text{\,erg\,cm$^{-3}$}} 
\newcommand{\gcc}{\text{\,g\,cm$^{-3}$}}   
\newcommand{\cs}{\text{\,$c^2$}}           
\newcommand{\pa}{\text{\,cm$^{1/2}$\,g$^{1/2}$\,s$^{-2}$}} 

\begin{document}
\title{Universal relations for elastic hybrid stars and quark stars}
\date{\today}

\author{Chun-Ming Yip\,\orcidlink{0000-0002-3311-5387}}
\affiliation{Department of Physics and Institute of Theoretical Physics, The Chinese University of Hong Kong, Shatin, New Territories, Hong Kong SAR, People’s Republic of China}
\affiliation{GSI Helmholtzzentrum f\"ur Schwerionenforschung, Planckstra{\ss}e 1, D-64291 Darmstadt, Germany}

\author{Shu Yan Lau\,\orcidlink{0000-0002-8239-0174}}
\affiliation{eXtreme Gravity Institute, Department of Physics, Montana State University, Bozeman, Montana 59717, USA}

\author{Kent Yagi\,\orcidlink{0000-0002-0642-5363}}
\affiliation{Department of Physics, University of Virginia, Charlottesville, Virginia 22904, USA}

\begin{abstract}
Some compact stars may contain deconfined quark matter, forming hybrid stars or quark stars.  
If the quark matter forms an inhomogeneous condensate in the crystalline color superconducting phase, its rigidity may be high enough to noticeably alter the stellar properties. 
In this paper, we investigate whether these elastic stars follow the universal relations, i.e., relations insensitive to equations of state, that have been well established for fluid stars. 
We improve upon previous studies by allowing quark matter in the background, static, and spherically symmetric configuration to be sheared.
Such background shear can be treated in the form of an effective pressure anisotropy. 
We then calculate the moment of inertia $I$, tidal deformability ${\lambda}_2$, and spin-induced quadrupole moment $Q$ of these models with pressure anisotropy. 
The $I$-${\lambda}_2$-$Q$ universal relations for the elastic hybrid (quark) star models are valid up to a variation of $\approx3\%(4\%)$, larger than that for typical fluid star models, when the maximal magnitude of quark matter shear modulus is considered in the crystalline color superconducting phase from realistic calculations.
The uncertainty in universal relations related to the stellar compactness for these elastic star models, on the other hand, remain comparable to those for typical fluid star models. 
Our results demonstrate the validity of universal relations for hybrid stars and quark stars with a realistic degree of pressure anisotropy due to the crystalline color superconducting quark matter.
\end{abstract}
\maketitle


\section{Introduction}
\label{sec: introduction}

The equation of state (EOS) for ultradense matter that constitutes compact stars is highly uncertain. 
Conventionally, compact stars are modeled using nuclear matter (NM) EOS, and are referred to as neutron stars (NSs). 
Alternatively, some compact stars may be hybrid stars (HSs) that contain a deconfined quark matter (QM) core inside the NM envelope~\cite{Alford:2004pf,Alford:2013aca}, or even quark stars (QSs) consisting of pure QM~\cite{Witten:1984rs}.
Although NSs are typically described as isotropic perfect NM fluids, the QM core of HSs and QSs may have an extremely high rigidity. 
For example, QM may form an inhomogeneous condensate in the crystalline color superconducting (CCS) phase~\cite{Alford:1997zt,Alford:1998mk,Alford:2000ze,Rajagopal:2006ig,Anglani:2013gfu}. 
After a core-collapse supernova explosion, the internal temperature of the newly formed protoneutron star rapidly drops from $\sim 10${\mev} to $\sim 0.1${\mev} within a few minutes through neutrino cooling. 
The formation of different quark phases inside the protoneutron star, including the CCS phase, was discussed in literature~\cite{Alford:2001dt,Alford:2007xm,Pagliara:2010na}.
The shear modulus of the CCS QM in compact stars was predicted to be up to $1000$ times higher than that of conventional NS crust models~\cite{Mannarelli:2007bs}.

Stellar elasticity gives rise to pressure anisotropy~\cite{Karlovini:2002fc,Karlovini:2003xi}, i.e., the radial pressure being different from the tangential one.
Several phenomenological anisotropy models (see, e.g., Refs.~\cite{Bowers:1974tgi,Horvat:2010xf}) were proposed to describe pressure anisotropy, where the degree of anisotropy can be parametrized.
As the source of pressure anisotropy is not explicitly specified in these phenomenological models, they can be applied to describe pressure anisotropy from different physical origins, including the QM shear modulus aforementioned, strong magnetic field~\cite{Huang:2009ue,Ferrer:2010wz,Yazadjiev:2011ks,Becerra:2024wku}, pion condensation~\cite{Sawyer:1973fv,Dickhoff:1981jq,Migdal:1990vm}, dark matter~\cite{Moraes:2021lhh,Bramante:2023djs,10.1093/mnras/stad3562}, and dark energy~\cite{Ghezzi:2009ct,Das:2014swa}.
A new theory of self-gravitating anisotropic fluids was recently formulated~\cite{Cadogan:2024mcl}, first in Newtonian gravity based on liquid crystal~\cite{Cadogan:2024ohj} and then extended to general relativity~\cite{Cadogan:2024ywc}.
On the other hand, the theory of elasticity in general relativity was formulated in Refs.~\cite{Karlovini:2002fc,Karlovini:2003xi,Alho:2021sli}, and was applied to model elastic stars~\cite{Karlovini:2002fc,Alho:2021sli,Dong:2024lte}. 
Macroscopic properties of HSs and QSs, such as the mass-radius ($M$-$R$) relations, stellar compactness $C \equiv M/R$, and Tolman-Oppenheimer-Volkoff (TOV) limits (maximum-mass configurations), can be significantly altered due to elasticity~\cite{Alho:2021sli,Alho:2022bki,Dong:2024lte}.

The universal relations between $C$ and quantities related to multipole moments, such as the moment of inertia $I$, tidal deformability $\lambda_2$, and spin-induced quadrupole moment $Q$~\cite{Hartle:1967he,Hartle:1968si}, were identified for slowly rotating or tidally deformed NSs~\cite{1994ApJ...424..846R,Bejger:2002ty,Lattimer:2004nj,10.1093/mnras/stt858,Baubock:2013gna,Breu:2016ufb,Jiang:2020uvb}, in the sense that these relations are insensitive to particular EOS within the uncertainty of $\sim 10\%$~\cite{Yagi:2016bkt}.
In 2013, the $I$-${\lambda}_2$-$Q$ universality for NSs was discovered with a validity of $\sim1\%$~\cite{Yagi:2013bca,Yagi:2013awa}. 
Later on, the investigation on universal relations was further extended to rapidly rotating stars~\cite{Doneva:2013rha,Pappas:2013naa,Chakrabarti:2013tca,Yagi:2014bxa,Paschalidis:2017qmb,Largani:2021hjo,Papigkiotis:2023onn,Kruger:2023olj}. 
We refer the readers to Ref.~\cite{Yagi:2016bkt} for a review of the universal relations. 
While the above studies focus on fluid stars, the tidal deformation~\cite{Lau:2017qtz,Lau:2018mae,Gittins:2020mll,Dong:2024opz} 
of elastic stars have also been studied. 
For example, Ref.~\cite{Dong:2024opz} showed that the $C-\lambda_2$ relations for elastic HSs deviates from that for fluid NSs, where the background configuration of the elastic HS models is assumed to be unsheared. 
Likewise, the universal relations for anisotropic NSs were also found to be slightly less robust compared to those for isotropic NSs~\cite{Yagi:2015hda,Das:2022ell,Mohanty:2023hha,Pretel:2024pem}.
These studies, nevertheless, employ phenomenological anisotropy models~\cite{Bowers:1974tgi,Horvat:2010xf} that parametrize the degree of pressure anisotropy, so the dependence of the deviation in the universal relations on the physical origin of pressure anisotropy is uncertain. 
It is also unclear whether the deviation remains noticeable when realistic elastic stellar background configurations and anisotropy models are considered when going beyond spherical symmetry. 
A comprehensive study of universal relations for elastic stars based on realistic sheared background configurations is still missing.

In this paper, we study the $I$-${\lambda}_2$-$Q$-$C$ relations for elastic HS and QS models in the slow rotation or small tidal deformation approximation 
by self-consistently constructing the static, spherically symmetric background configuration of these elastic star models in general relativity~\cite{Karlovini:2002fc} under the influence of QM in the CCS phase~\cite{Mannarelli:2007bs}. 
As already mentioned, Karlovini and Samuelsson~\cite{Karlovini:2002fc} showed that background shear can be treated within the framework of effective pressure anisotropy.
Thus, we follow the work of Ref.~\cite{Yagi:2015hda} on universal relations for anisotropic NSs to construct slowly rotating or tidally deformed elastic star models. 
While the $I$-${\lambda}_2$-$Q$-$C$ relations have been demonstrated to be insensitive to EOS for fluid NSs, we aim to investigate their validity for the elastic stars with a realistic degree of pressure anisotropy associated with the shear modulus of QM in the CCS phase. 

Let us briefly summarize our main findings.
The $M$-$R$ relations and stellar compactness $C$ for elastic HSs and QSs can be significantly altered when the rigidity of QM in the CCS phase is taken into account.
The TOV limits of the elastic HSs and QSs are enhanced by a few percent with respect to their fluid limits in the extreme cases. 
For all elastic star models we examined, $C$ increases, but $\bar{I}$, $\bar{\lambda}_2$, and $\bar{Q}$ decrease relative to their fluid limits when comparing two configurations with the same central pressure. 
Systematic shifts in the $I$-${\lambda}_2$-$Q$-$C$ relations associated with the QM shear modulus are found for these elastic star models.
The $I$-${\lambda}_2$-$Q$ relations for the elastic HS (QS) models can show an EOS variation of $\approx3\%(4\%)$, more pronounced than typical fluid star models.
Meanwhile, the deviation in the universal relations involving $C$ for these elastic stars is still compatible with those for other fluid star models.

The remainder of the paper is organized as follows.
In Sec.~\ref{sec: methods}, we introduce the elastic HS and QS models used in this paper.
In Sec.~\ref{sec: results}, we present our numerical results of the $I$-${\lambda}_2$-$Q$-$C$ relations for these models and discuss the impacts of the QM shear modulus on the universal relations.
Finally, we summarize our findings and compare the results with other relevant literature in Sec.~\ref{sec: conclusions and discussions}. 
We use the geometrized unit system with $G = c = 1$ throughout the paper unless otherwise specified.

\section{Methods}
\label{sec: methods}

In this paper, we use the quasi-Hookean EOS to describe the total energy density $\rho$ of elastic materials, 
\begin{equation}
    \rho = \tilde{\rho} + \tilde{\mu} S^2,
    \label{eq: quasi-Hookean EOS}
\end{equation}
where $\tilde{\mu}$ is the shear modulus and $S^2$ is the shear scalar. 
Equation~\eqref{eq: quasi-Hookean EOS} expresses $\rho$ as a sum of the unsheared energy density $\tilde{\rho}$ and a sheared component depending on the magnitude of shear deformation. 
The unsheared pressure $\tilde{p}$ is related to $\tilde{\rho}$ by further providing a fluid EOS.
We refer the readers to Refs.~\cite{Karlovini:2002fc,Dong:2024lte} for more details on the quasi-Hookean EOS. 

To construct HS models, we need hybrid EOS to relate $\tilde{p}$ and $\tilde{\rho}$ for the QM core and the NM envelope.
We employ the constant speed of sound (CSS) parametrization~\cite{Alford:2013aca} to construct the hybrid EOS via the Maxwell construction, 
\begin{equation}
    \tilde{\rho}(\tilde{p}) = 
    \begin{cases}
        \rho_\text{NM}(\tilde{p})\,, & (\tilde{p} \leq p_\text{trans}), \\
        \rho_\text{NM}(p_\text{trans}) + \Delta\rho + c_\text{QM}^{-2} \left( \tilde{p} - p_\text{trans} \right)\,,& (\tilde{p} > p_\text{trans}).
    \end{cases}
    \label{eq: CSS template}
\end{equation}
Here, the transition pressure $p_\text{trans}$, energy gap $\Delta\rho$, and QM sound speed squared $c_\text{QM}^2$ are the CSS parameters that determine the properties of the QM part of the hybrid EOS. 
A NM EOS that expresses the NM energy density $\rho_\text{NM}$ as a function of $\tilde{p}$ up to $p_\text{trans}$ is also required.
We use the  Akmal-Pandharipande-Ravenhall (APR) EOS\footnote{\href{https://compose.obspm.fr/EOS/328}{https://compose.obspm.fr/EOS/328}}~\cite{Akmal:1998cf,Haensel:2007yy,Davis:2024nda,Davis:2025nwz} to construct the NM part of the hybrid EOS. 
Alternatively, we use the Bombaci-Logoteta (BL) EOS\footnote{\href{https://compose.obspm.fr/EOS/121}{https://compose.obspm.fr/EOS/121}}~\cite{Douchin:2001sv,Bombaci:2018ksa} and the Shen-Toki-Oyamatsu-Sumiyoshi (STOS) EOS\footnote{\href{https://compose.obspm.fr/EOS/71}{https://compose.obspm.fr/EOS/71}}~\cite{Shen:1998by,Shen:1998gq,Sugahara:1993wz}, at $\beta$-equilibrium and the lowest temperature available ($0.1${\mev}), for constructing other hybrid EOS in order to investigate the effects of varying the NM EOS on our main results. 
These EOS tables are available in CompOSE~\cite{Typel:2013rza,Oertel:2016bki,CompOSECoreTeam:2022ddl}. 
Meanwhile, QS models can also be constructed using Eq.~\eqref{eq: CSS template} by neglecting the NM part.

By computing the low-energy effective Lagrangian for the phonon modes that originate from the spontaneous breaking of translation invariance by the crystalline condensates, Ref.~\cite{Mannarelli:2007bs} derived the shear modulus of QM in the CCS phase, 
\begin{equation}
    \tilde{\mu} = 2.47 \, \text{MeV\,fm$^{-3}$} \left( \frac{\Delta}{10 \,\text{MeV}} \right)^2 \left( \frac{\mu_q}{400 \,\text{MeV}} \right)^2,
    \label{eq: QM shear modulus analytical}
\end{equation}
where $\Delta$ is the gap parameter between about $5$ and $25$ {\mev} and $\mu_q$ is the quark chemical potential. 
At high-density limit, $\rho \propto \mu_q^4$ for ultrarelativistic free Fermi gas, and hence the above equation reduces to a form of~\cite{Faggert:2023kvm,Dong:2024lte}
\begin{equation}
    \tilde{\mu} = \kappa \sqrt{\rho},
    \label{eq: QM shear modulus}
\end{equation}
where $\kappa$ is a fitting coefficient to Eq.~\eqref{eq: QM shear modulus analytical}. 
In the following calculations, we adopt Eq.~\eqref{eq: QM shear modulus} to approximate the CCS QM shear modulus inside the HSs and QSs. 
The maximum value of $\kappa$ is $7\times10^{26}${\pa} when taking $\Delta = 25$ {\mev} in Eq.~\eqref{eq: QM shear modulus analytical}.

The static, spherically symmetric background configuration of the elastic HS and QS models is modeled with anisotropic pressure. 
The stress-energy tensor for elastic stars with scalar pressure anisotropy is given by~\cite{Yagi:2015hda}
\begin{equation}
    T_{\alpha\beta} = \rho u_\alpha u_\beta + p_r k_\alpha k_\beta + p_t \Pi_{\alpha\beta},
    \label{eq: stress-energy tensor}
\end{equation}
where $p_r$ ($p_t$) is the radial (tangential) pressure, $u^\alpha$ is the four-velocity, $k^\alpha$ is the unit normal vector in the radial direction orthogonal to $u^\alpha$, and $\Pi_{\alpha\beta}$ is the projection operator onto a two-surface orthogonal to both $u^\alpha$ and $k^\alpha$, 
\begin{equation}
    \Pi_{\alpha\beta} = g_{\alpha\beta} + u_\alpha u_\beta - k_\alpha k_\beta,
\end{equation}
where $g_{\alpha\beta}$ is the spacetime metric. 
We calculate the stellar compactness $C$ of the stars from this background configuration. 
We refer the readers to Appendix~\ref{app: background and perturbation equations} for the derivation of the corresponding background equations.

The moment of inertia $I$, tidal deformability $\lambda_2$, and spin-induced quadrupole moment $Q$ of the elastic HS and QS models are obtained by imposing perturbations to the metric up to the second order in slow rotation or the first order in small tidal deformation.
Once again, we refer the readers to Appendix~\ref{app: background and perturbation equations} for the details of the corresponding perturbation equations.
We next define the following dimensionless quantities:
\begin{equation}
    \Bar{I} \equiv \frac{I}{M^3},\, \Bar{\lambda}_2 \equiv \frac{\lambda_2}{M^5},\, \Bar{Q} \equiv \frac{Q}{M^3 \chi^2},
\end{equation}
where $\chi \equiv J / M^2$ with $J$ being the magnitude of the spin angular momentum, and we study their universal relations.

Based on astrophysical observations and terrestrial experiments, all the HS and QS models presented in this paper, unless otherwise specified, are compatible with the following constraints:
\begin{enumerate}

    \item The maximum wave speeds do not violate the causality limit (see Refs.~\cite{Karlovini:2002fc,Dong:2024lte} for the calculation of different wave speeds in elastic materials).
    
    \item The $M$-$R$ relations are compatible with the gravitational-wave event {GW170817} from a binary NS merger~\cite{LIGOScientific:2018cki}, as well as the measurement of typical-mass ($\approx1.4${\msol}) and high-mass ($\approx2${\msol}) pulsars by NICER~\cite{Miller:2019cac,Salmi:2024aum}.
    
    \item For the HS models, the transition densities on the NM side are above the nuclear saturation density $\rho_0 \equiv 2.6\times10^{14}${\gcc}.

    \item For the QS models, the $M$-$R$ relations are compatible with the observation of the central compact object within the supernova remnant {HESS J1731-347}~\cite{Doroshenko:2022nwp}, which is hypothesized to be a QS~\cite{Rather:2023tly,DiClemente:2022wqp}.
    
\end{enumerate}

Note that other constraints on the EOS properties were proposed. 
For example, the pressure at twice the nuclear saturation density is constrained to be $3.5^{+2.7}_{-1.7} \times 10^{34}${\ecc}, at the $90\%$ credible interval, based on the analysis of the gravitational-wave event {GW170817}~\cite{LIGOScientific:2018cki}. 
Reference~\cite{Alford:2013aca} applied the Seidov stability condition~\cite{1971SvA....15..347S} to define the critical energy gap $\Delta\rho_\text{crit}$, 
\begin{equation}
    \frac{\Delta\rho_\text{crit}}{\rho_\text{trans}}
    = \frac{1}{2} + \frac{3}{2} \frac{p_\text{trans}}{\rho_\text{trans}},
    \label{eq: Seidov stability condition}
\end{equation}
where $\rho_\text{trans}$ is the energy density at $p_\text{trans}$ for the NM EOS. 
When $\Delta\rho > \Delta\rho_\text{crit}$, a cusp or even a disconnected branch emerges at the phase transition point in the $M$-$R$ relations for HSs (see Ref.~\cite{Alford:2013aca} for the detailed discussions on the possible topologies of the $M$-$R$ relations for HSs).
To explore the variation in universal relations for elastic HSs and QSs within regions in the parameter space close to the boundaries of the allowed region, we do not impose these EOS property constraints explicitly on our HS and QS models. 

We first construct a canonical HS model with a specific set of CSS parameters $\{ p_\text{trans}, \Delta\rho, c_\text{QM}^2 \}$ as an example to illustrate the general impacts of the QM shear modulus on the elastic stars. 
After that, we scan through the CSS parameter space to examine the validity of universal relations that have been established for fluid NSs with respect to these HS models. 
We also discuss the impact of using different NM EOS to construct the HS models on the results.
Finally, we perform the same on the elastic QS models.

\section{Results}
\label{sec: results}

We now present, in turn, our results for the HS and QS models.

\subsection{Hybrid stars}

\subsubsection{Canonical model}

\begin{figure}
    \includegraphics[width=0.95\columnwidth]{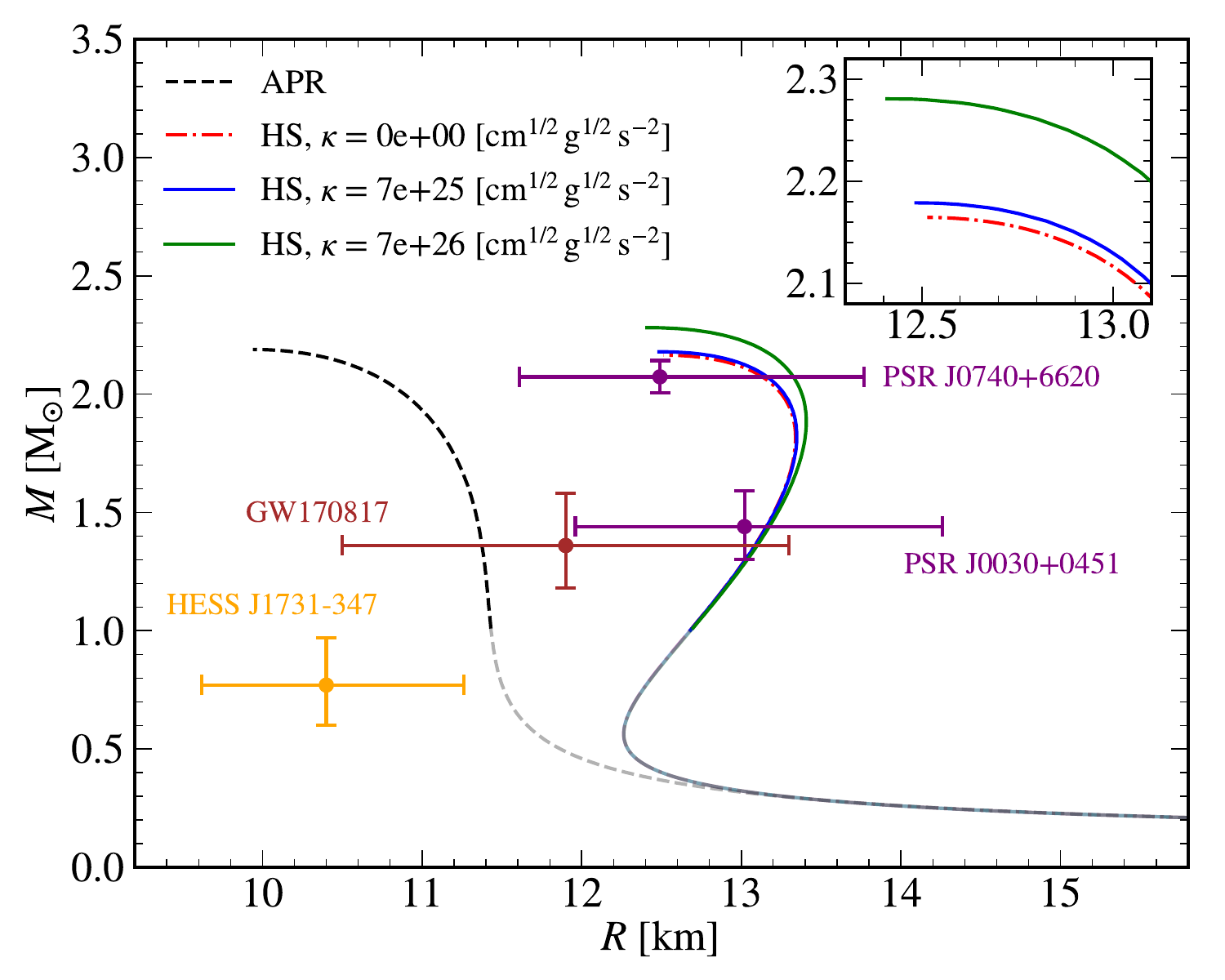}
    \caption{
    $M$-$R$ relations for the {APR} EOS model (black dashed curve) and the canonical HS model (see the main text for the definition) with different values of $\kappa$. 
    The red dash-dotted curve represents the fluid limit, and the blue (green) solid curves represent the elastic models with $\kappa = 7\times10^{25(26)}${\pa}. 
    These curves transit from translucent to opaque at $1${\msol}, and terminate at their TOV limits. 
    The HS models satisfy the $M$-$R$ relations for the gravitational-wave event {GW170817} (brown error bar) at the $90\%$ credible interval~\cite{LIGOScientific:2018cki}, as well as the pulsars {PSR J0030+0451} (an updated constraint from this pulsar is also available~\cite{Vinciguerra:2023qxq}) and {PSR J0740+6620} (purple error bars) measured by NICER at the $68\%$ credible intervals~\cite{Miller:2019cac,Salmi:2024aum}. 
    We also present the $M$-$R$ relation for the central compact object within the supernova remnant {HESS J1731-347} (orange error bar) at the $68\%$ credible interval~\cite{Doroshenko:2022nwp} as a reference.
    The subplot in the upper-right corner focuses on the $M$-$R$ relations for the HS models near the TOV limits.
    }
    \label{fig: M-R, HS, APR, vary kappa}
\end{figure}

\begin{figure}
    \includegraphics[width=0.95\columnwidth]{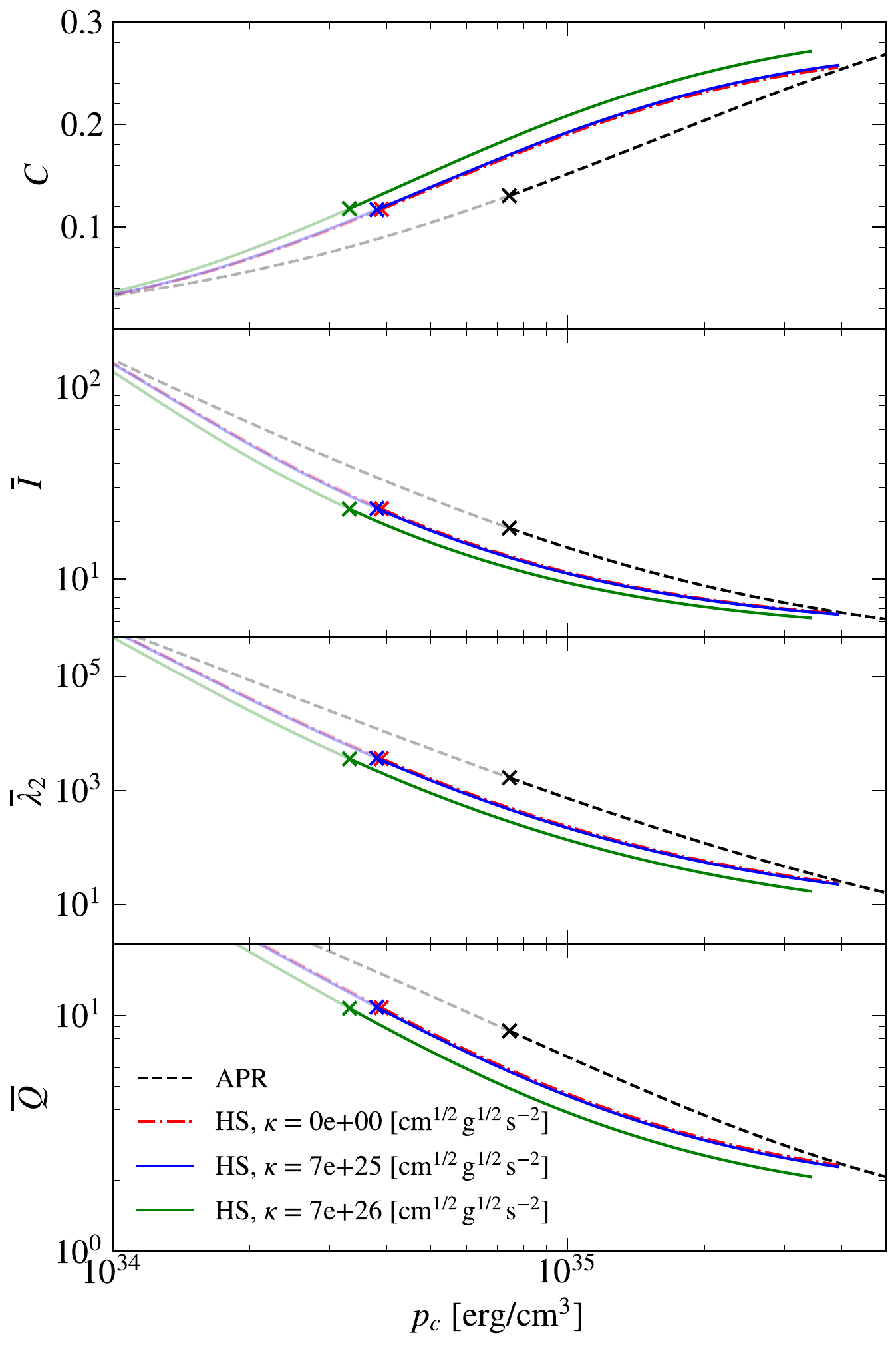}
    \caption{$C$, $\bar{I}$, $\bar{\lambda}_2$, and $\bar{Q}$ versus $p_c$ for the models displayed in Fig.~\ref{fig: M-R, HS, APR, vary kappa}.
    The configuration at $1${\msol} is indicated by a marker in each curve.
    All the HS models above $1${\msol} in the plots have a sufficiently high value of $p_c$ to contain a QM core. 
    The elastic models always have larger $C$ but smaller $\bar{I}$, $\bar{\lambda}_2$, and $\bar{Q}$ relative to the fluid limit at the same $p_c$.
    }
    \label{fig: vs_pc-4, HS, APR, vary kappa}
\end{figure}

\begin{figure*}
    \subfigure[$I$-${\lambda}_2$ relations]
    {\includegraphics[width=0.95\columnwidth]{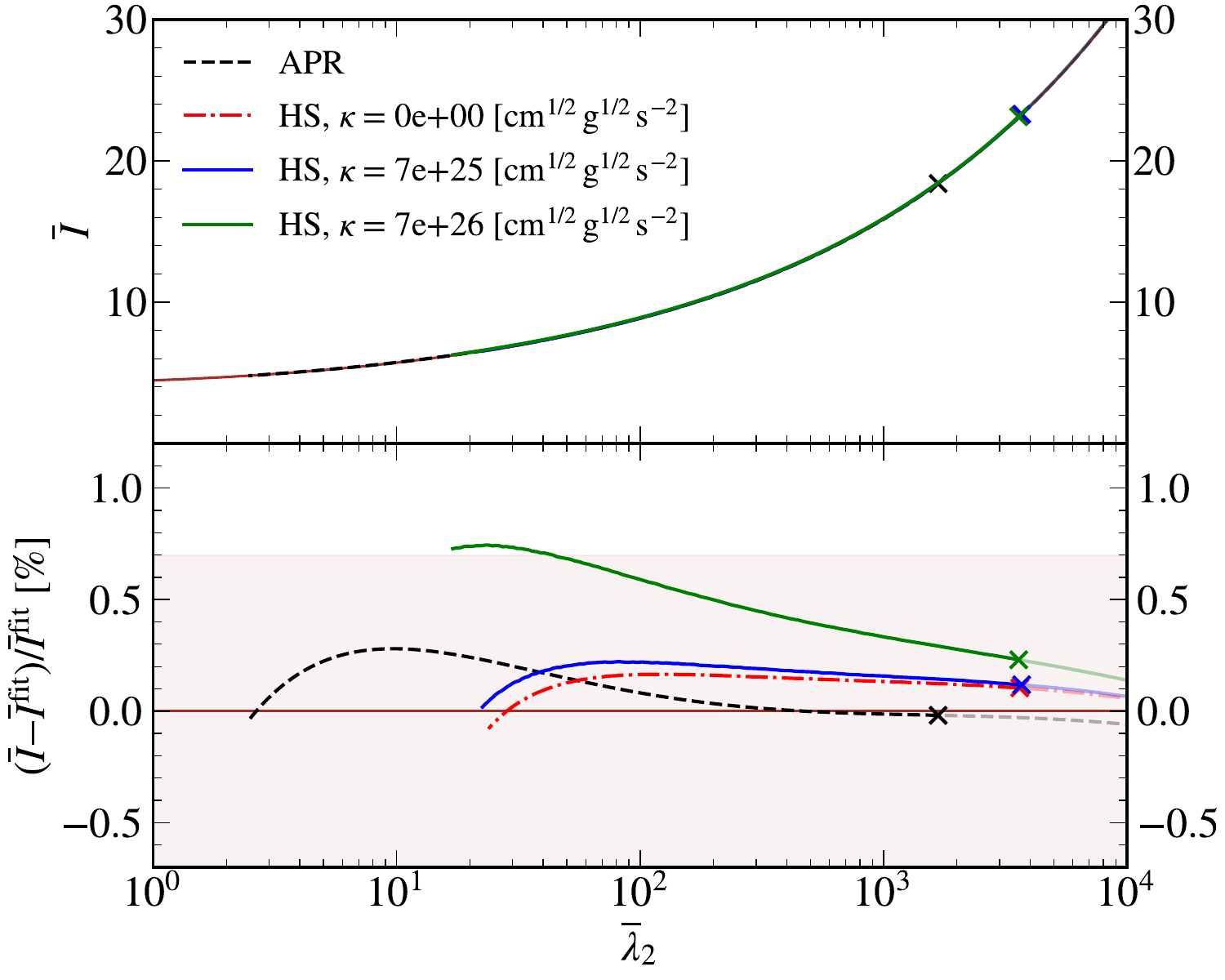}}
    \subfigure[$I$-$Q$ relations]
    {\includegraphics[width=0.95\columnwidth]{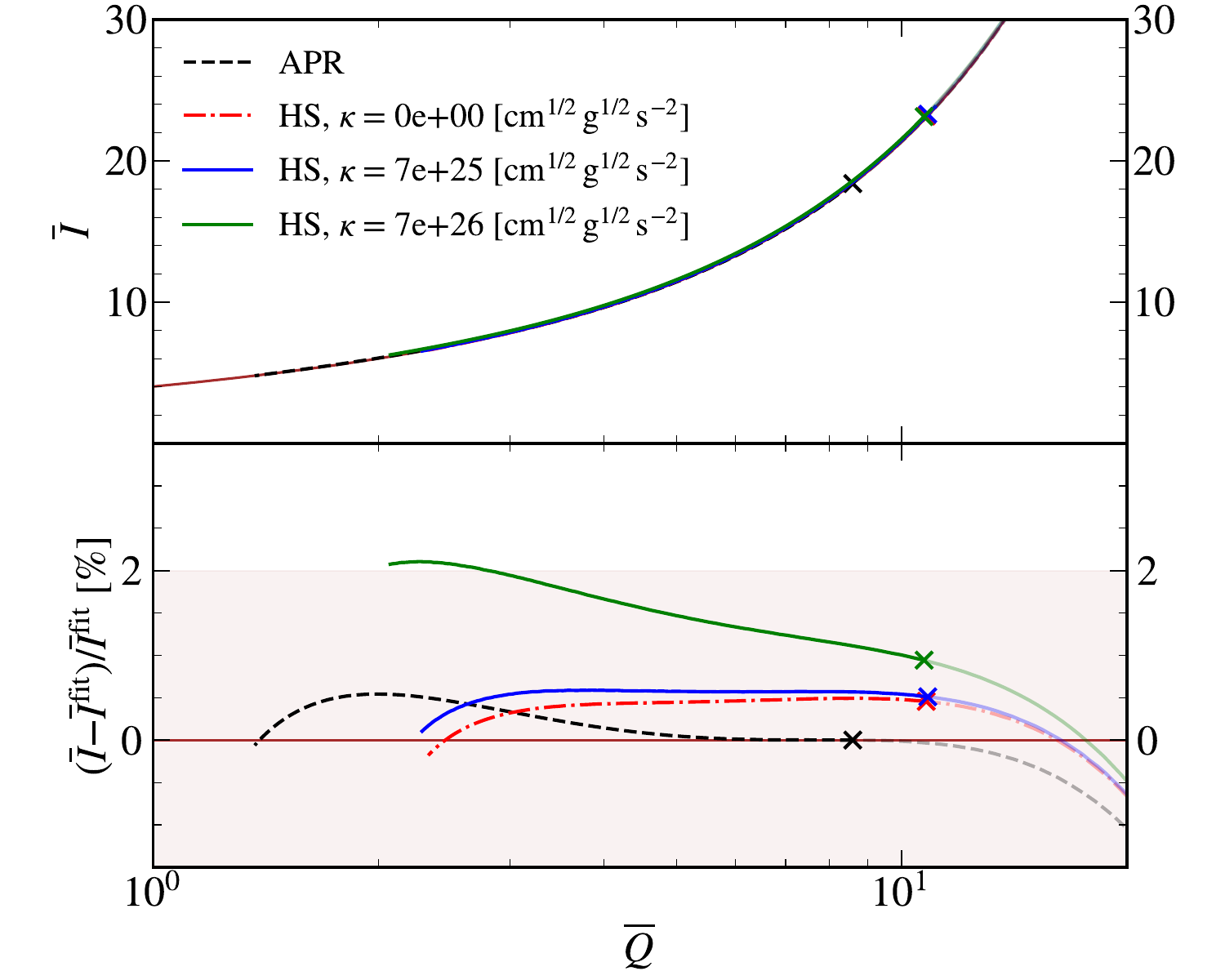}\label{fig: uni, HS, APR, vary kappa, I-Q}}
    \subfigure[$Q$-${\lambda}_2$ relations]
    {\includegraphics[width=0.95\columnwidth]{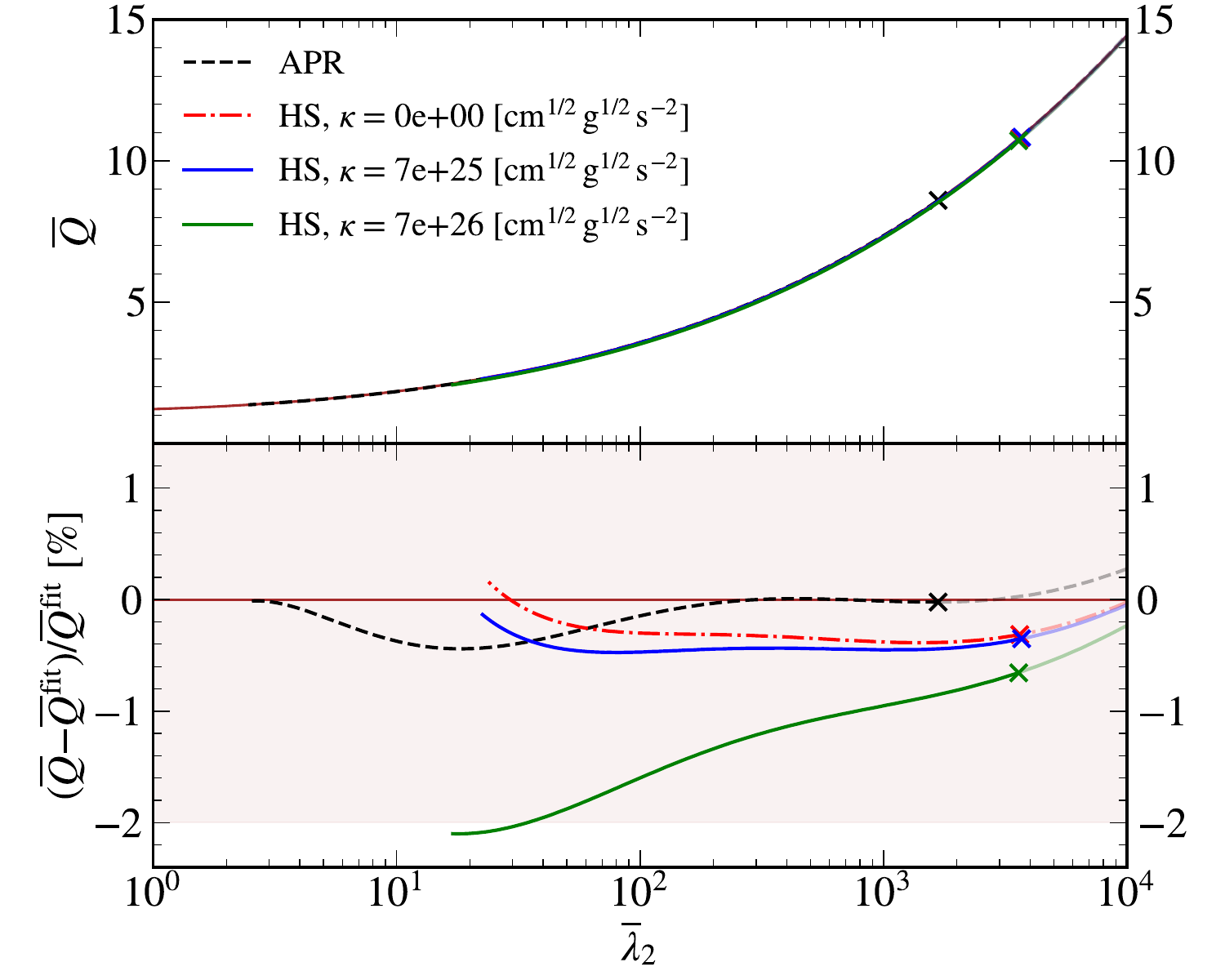}}
    \subfigure[$I$-$C$ relations]
    {\includegraphics[width=0.95\columnwidth]{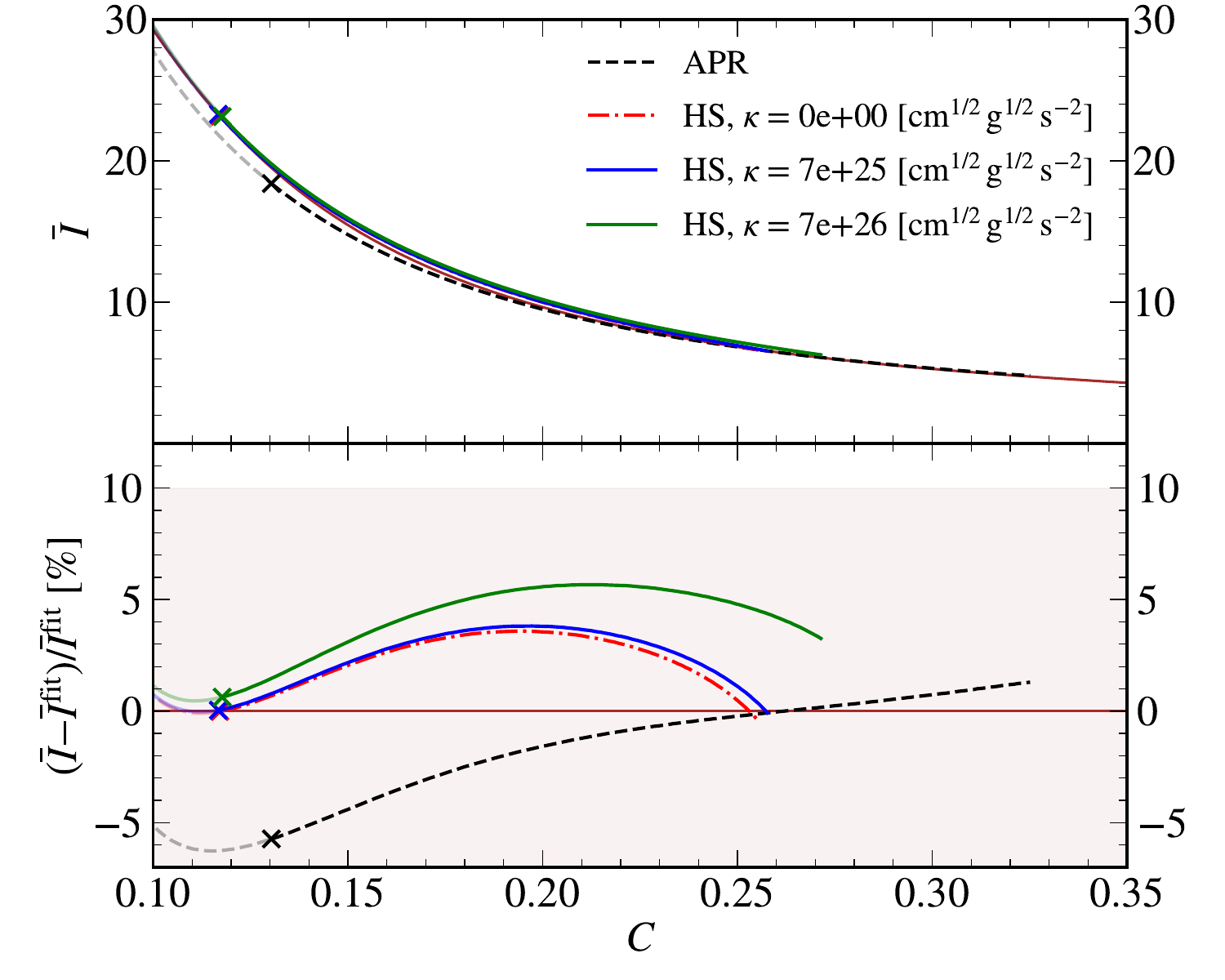}\label{fig: uni, HS, APR, vary kappa, I-C}}
    \subfigure[$C$-${\lambda}_2$ relations]
    {\includegraphics[width=0.95\columnwidth]{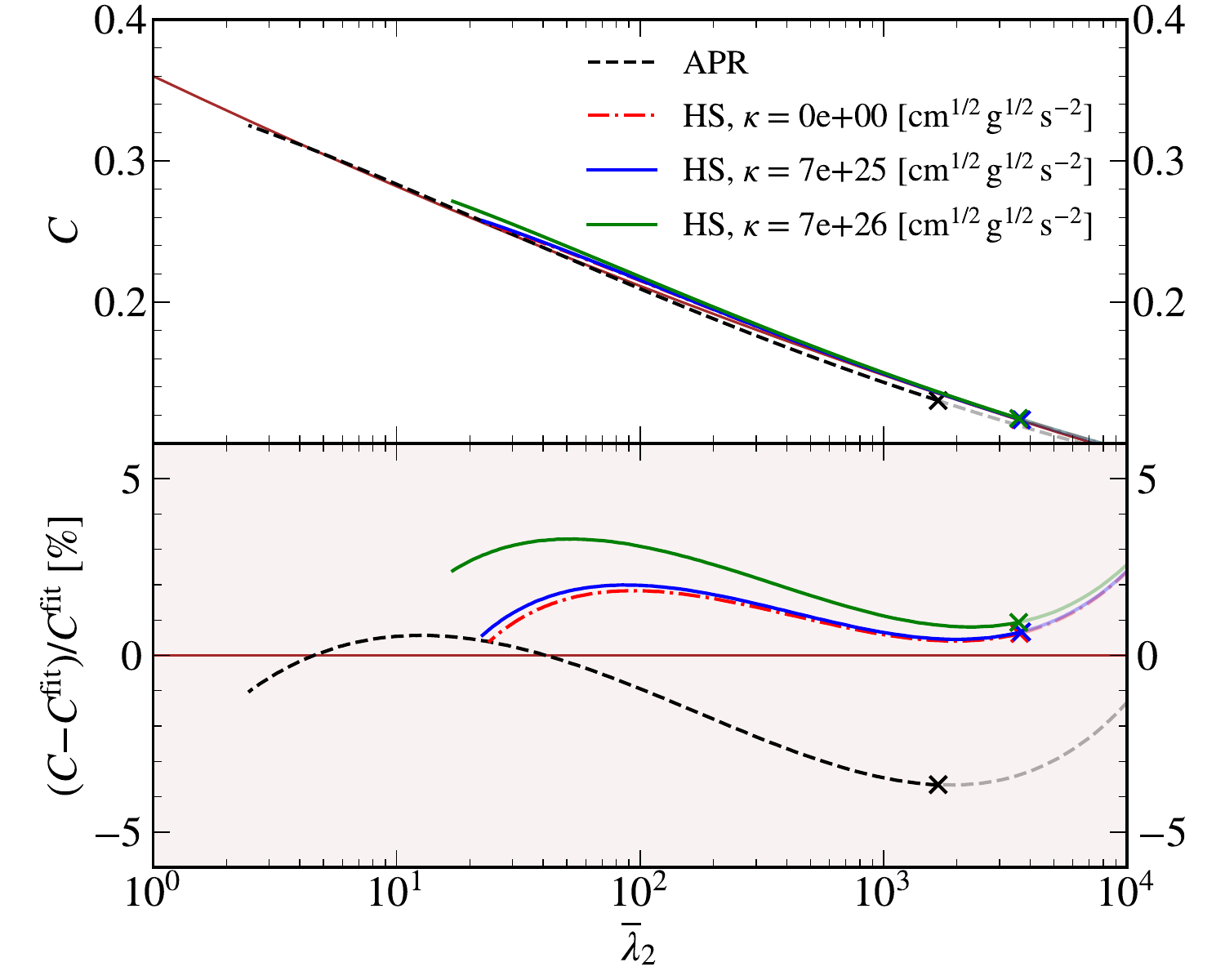}}
    \subfigure[$Q$-$C$ relations]
    {\includegraphics[width=0.95\columnwidth]{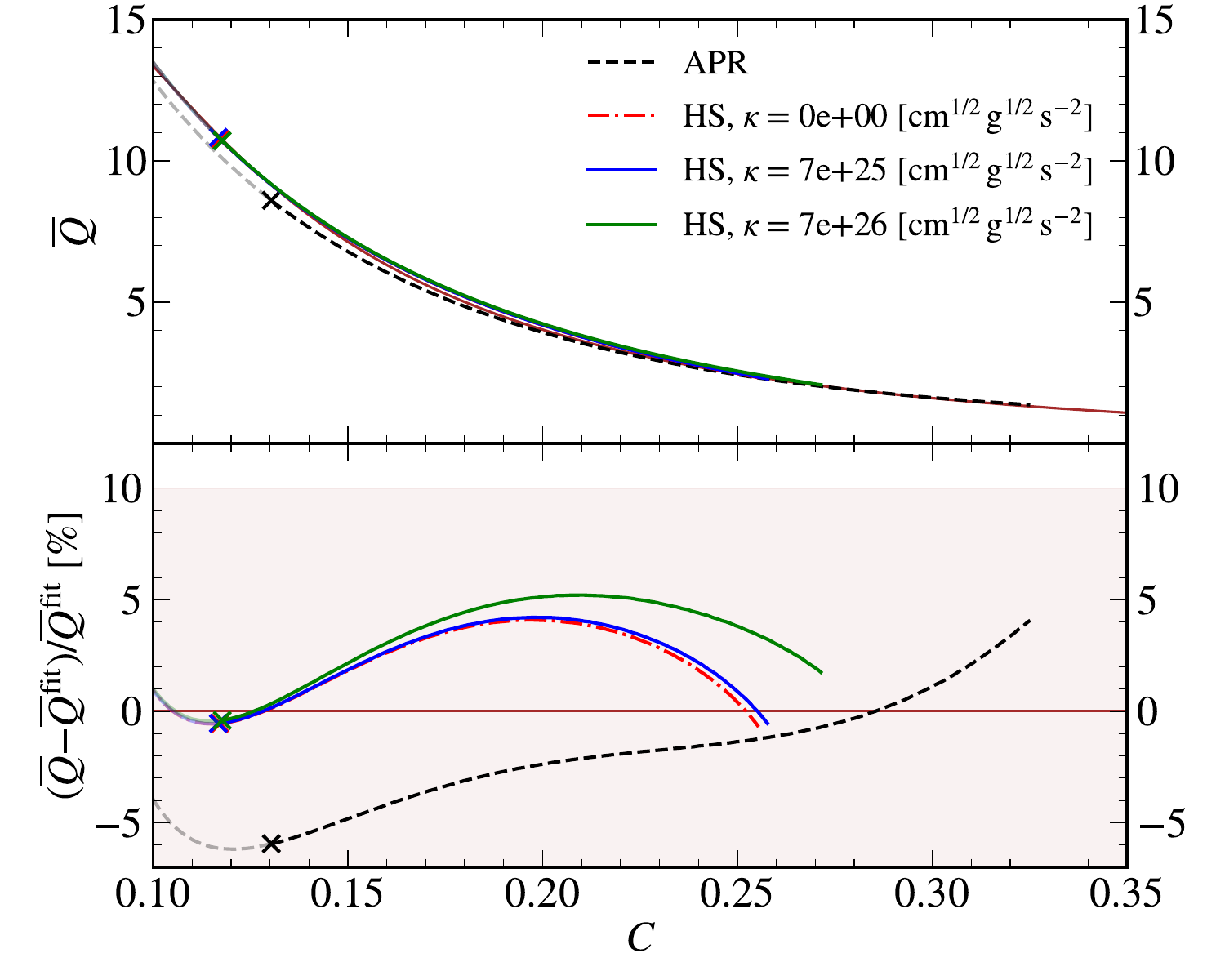}}
    \caption{
    Upper: universal relations for the models displayed in Fig.~\ref{fig: M-R, HS, APR, vary kappa}. 
    Lower: percentage relative difference in the universal relations between the models and the fitting formulas~\cite{Yagi:2016bkt}.  
    The brown solid curves represent the fitting formulas.
    The maximum variability of the fitting formulas for fluid NSs is further shaded in the lower panels: (a) $I$-${\lambda}_2$ relations, $0.7\%$; (b) $I$-$Q$ relations, $2\%$; (c) $Q$-${\lambda}_2$ relations, $2\%$; (d) $I$-$C$ relations, $10\%$; (e) $C$-${\lambda}_2$ relations: $7\%$; (f) $Q$-$C$ relations: $10\%$.
    Unlike most of the other literature, we do not take the absolute value of the percentage relative difference in the lower panels, so that the systematic shifts due to the QM shear modulus can be read off more easily.
    }
    \label{fig: uni, HS, APR, vary kappa}
\end{figure*}

\begin{figure*}[!ht]
    \subfigure[$M$-$R$ relations]{\includegraphics[width=0.95\columnwidth]{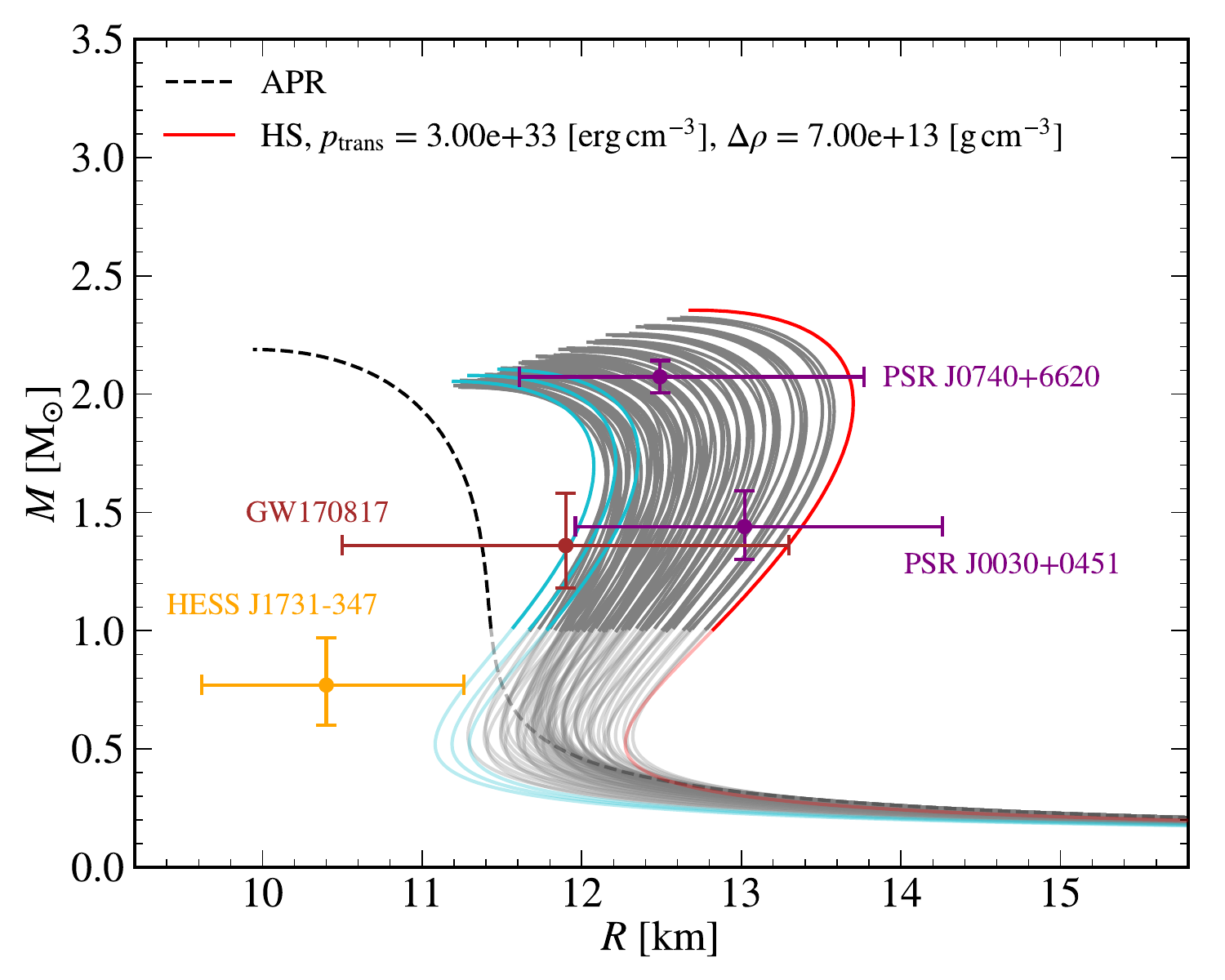}\label{fig: 2d, HS, APR, v=033, M-R}}
    \subfigure[$I$-$Q$ relations]{\includegraphics[width=0.95\columnwidth]{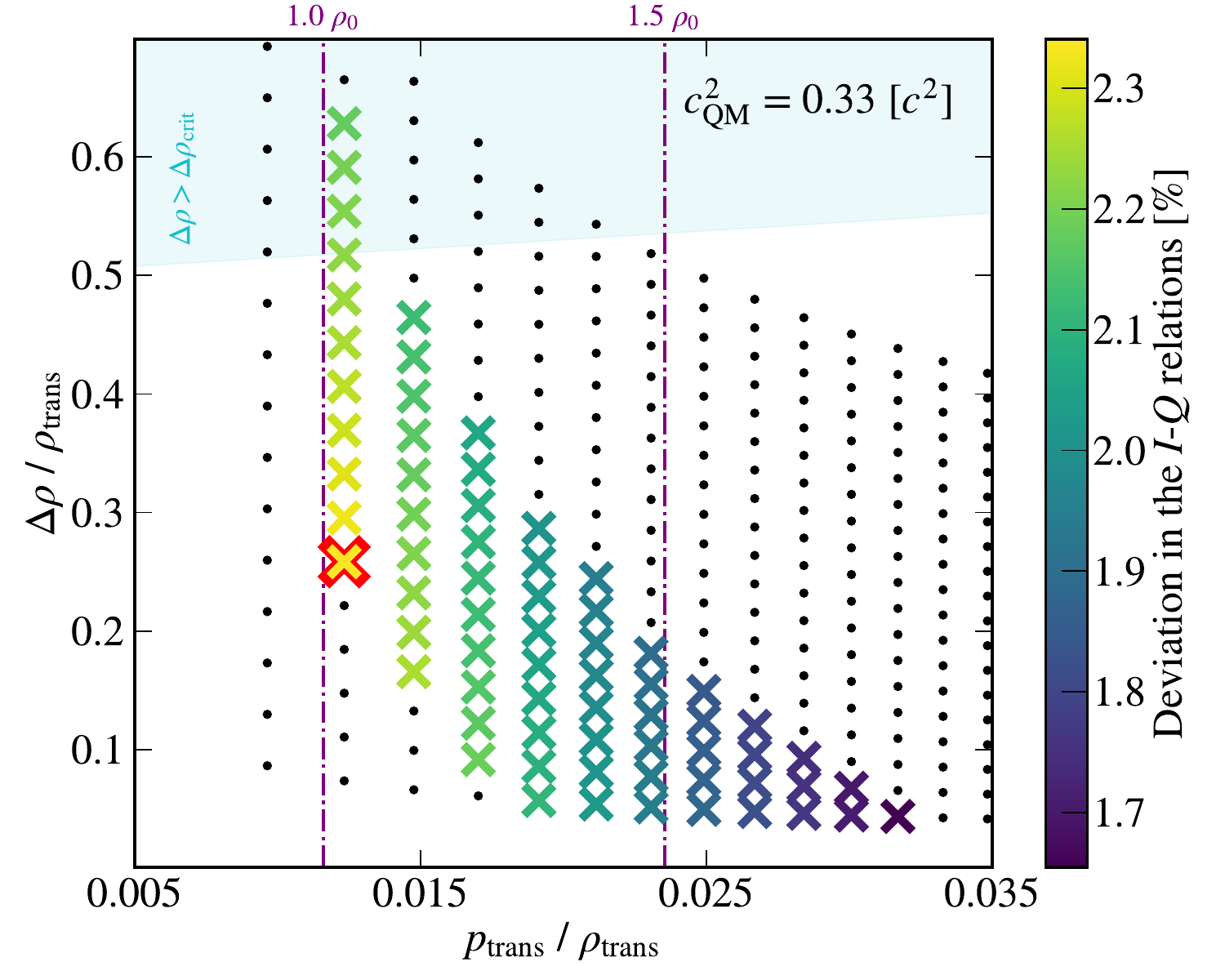}\label{fig: 2d, HS, APR, v=033, I-Q}}
    \caption{
    (a) Same as Fig.~\ref{fig: M-R, HS, APR, vary kappa}, but for a collection of HS models with varying values of $p_\text{trans}$ and $\Delta\rho$. 
    These models adopt $c_\text{QM}^2 = 0.33{\cs}$ and $\kappa=7\times10^{26}${\pa}.
    The HS models violating the Seidov stability condition, i.e., with $\Delta\rho > \Delta\rho_\text{crit}$, are shown in cyan, and the remaining are shown in gray.
    (b) Deviation in the $I$-$Q$ relations, which represent the extrema of percentage relative difference between these HS models and the fitting formulas~\cite{Yagi:2016bkt}, in $p_\text{trans} / \rho_\text{trans}$-$\Delta\rho / \rho_\text{trans}$ plane. 
    To demonstrate the parameter space explored, we also use black dots to indicate the models that are not fully compatible with all the constraints listed in Sec.~\ref{sec: methods}.
    The region where the Seidov stability condition is violated is shaded in cyan.
    The two purple dash-dotted vertical lines indicate the position where $\rho_\text{trans} = 1.0$ and $1.5 \rho_0$, respectively.
    In both panels, the particular model with the largest magnitude of deviation in the $I$-$Q$ relations is highlighted in red.
    }
    \label{fig: 2d, HS, APR, v=033}
\end{figure*}

\begin{figure*}[!ht]
    \subfigure[$M$-$R$ relations]{\includegraphics[width=0.95\columnwidth]{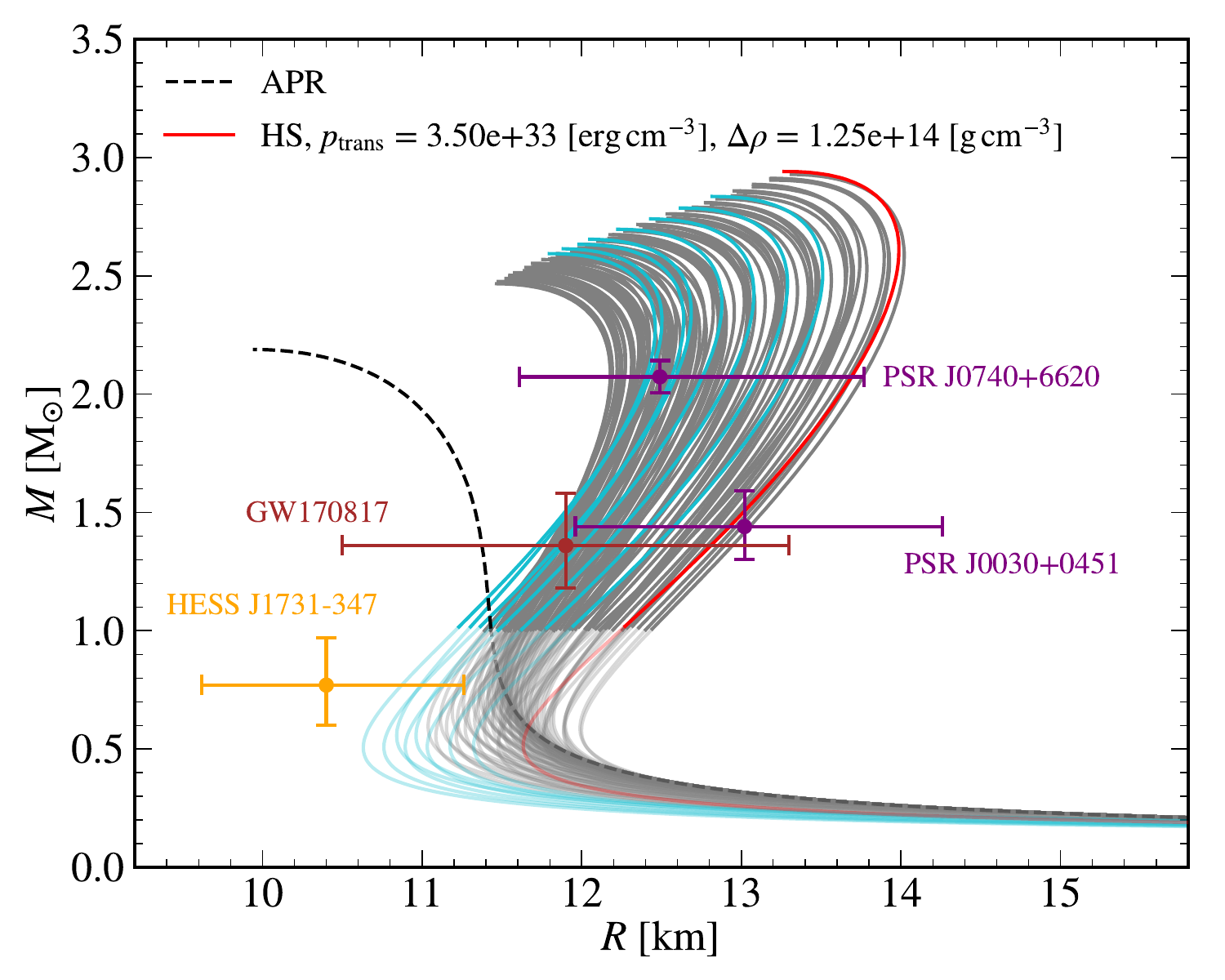}}
    \subfigure[$I$-$Q$ relations]{\includegraphics[width=0.95\columnwidth]{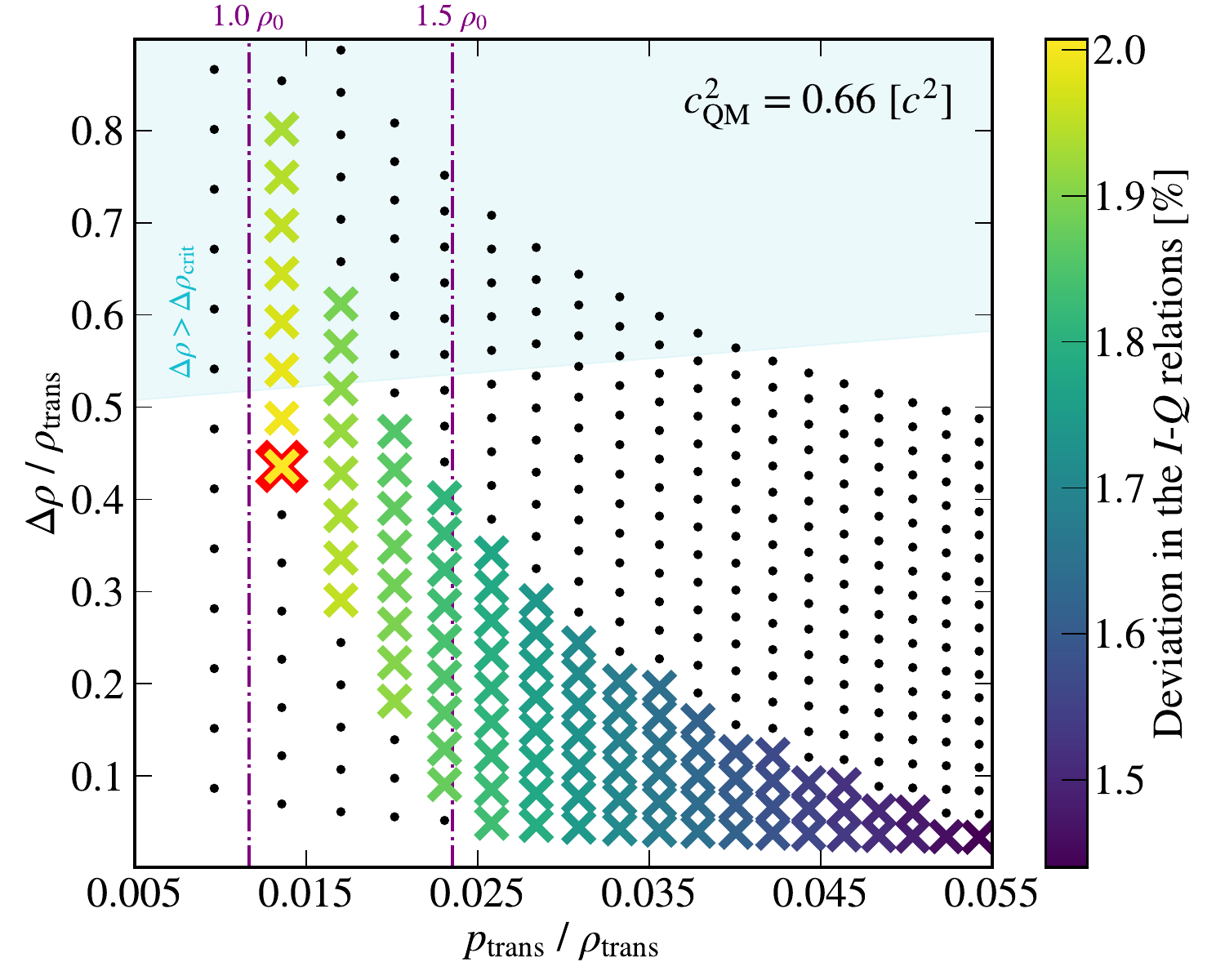}}
    \caption{
    Same as Fig.~\ref{fig: 2d, HS, APR, v=033}, but for HS models with $c_\text{QM}^2 = 0.66{\cs}$.
    }
    \label{fig: 2d, HS, APR, v=066}
\end{figure*}

\begin{figure*}[!ht]
    \subfigure[$M$-$R$ relations]{\includegraphics[width=0.95\columnwidth]{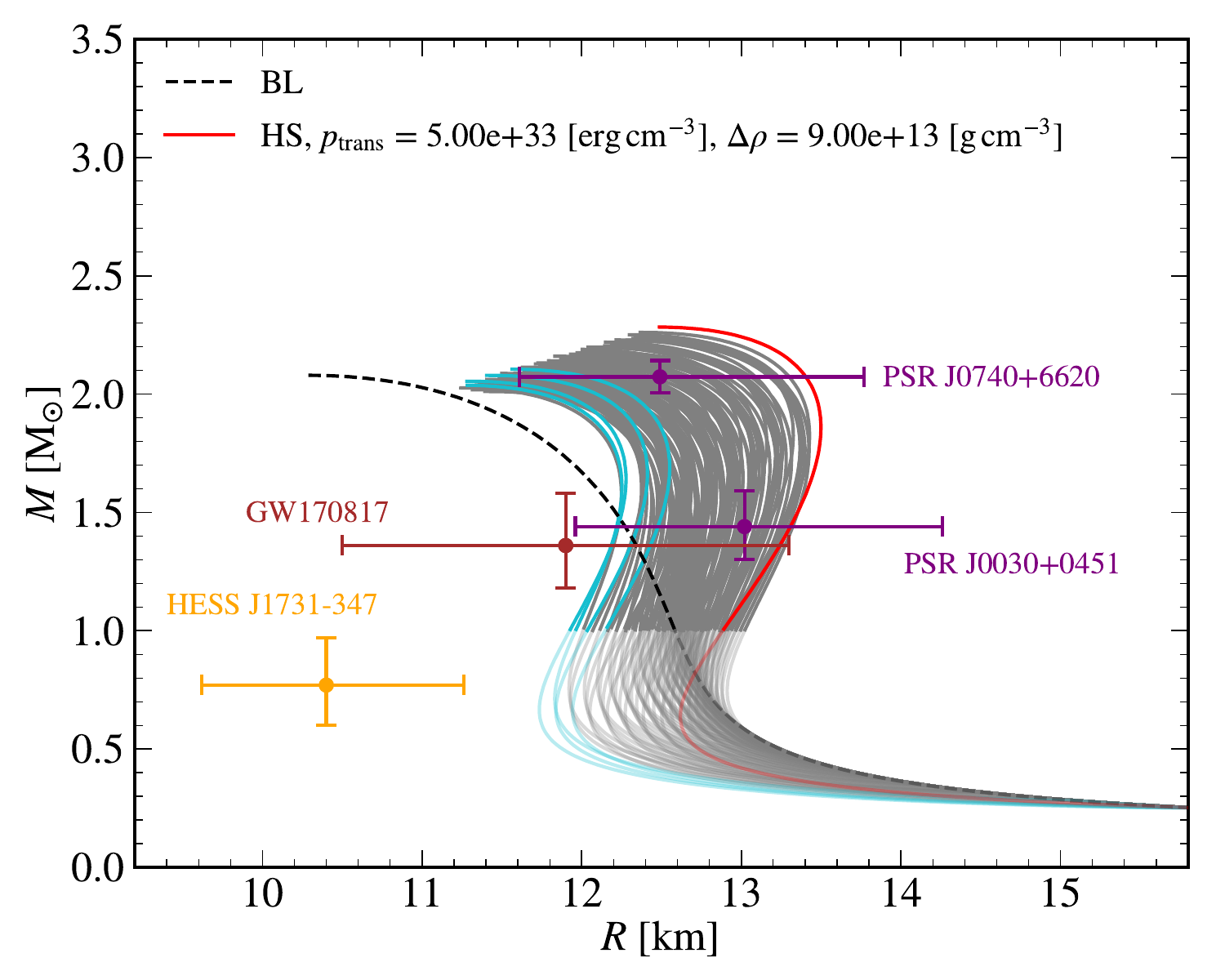}}
    \subfigure[$I$-$Q$ relations]{\includegraphics[width=0.95\columnwidth]{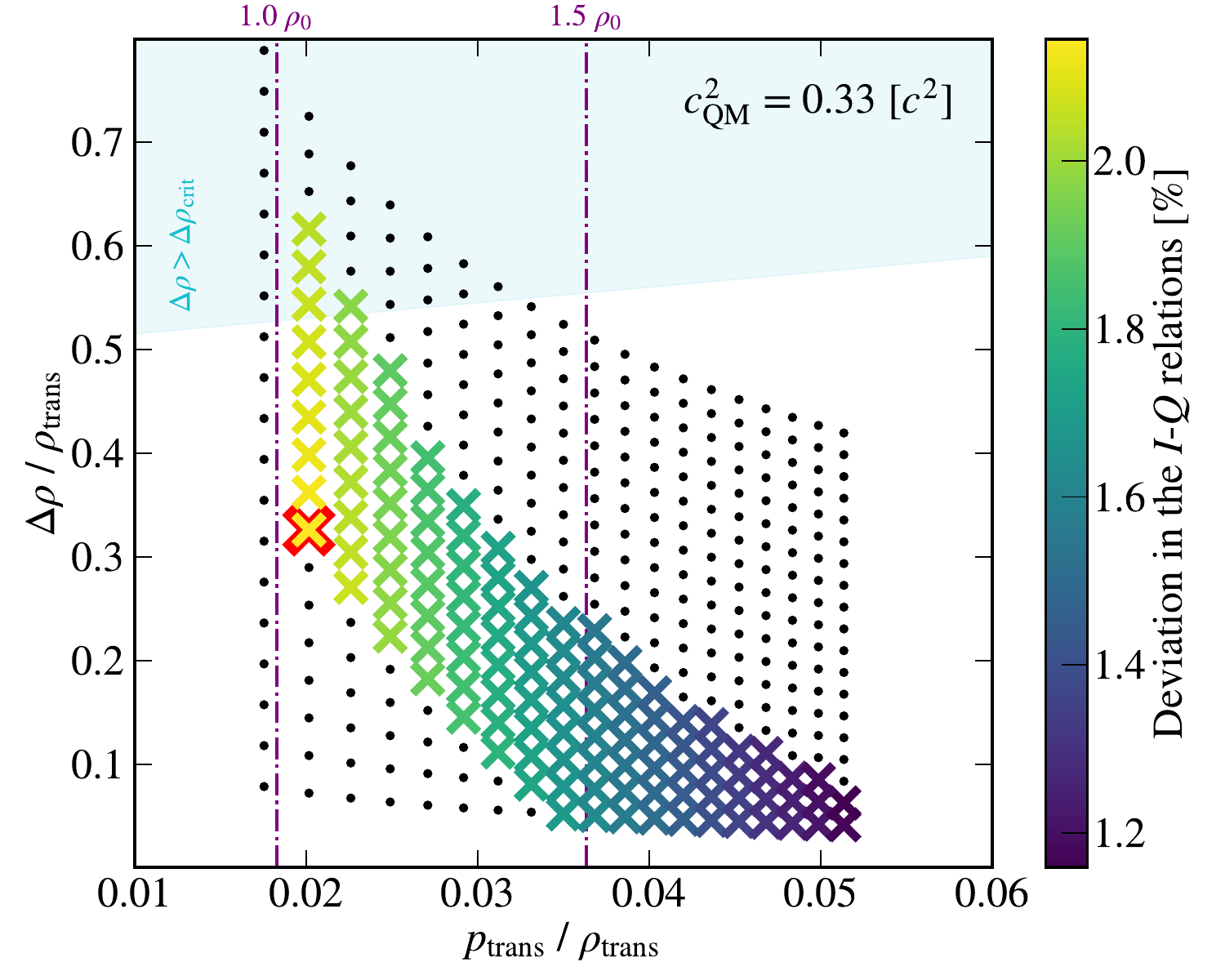}}
    \caption{
    Same as Fig.~\ref{fig: 2d, HS, APR, v=033}, but for HS models with the {BL} EOS being the NM part.
    }
    \label{fig: 2d, HS, BL, v=033}
\end{figure*}

\begin{figure*}[!ht]
    \subfigure[$M$-$R$ relations]{\includegraphics[width=0.95\columnwidth]{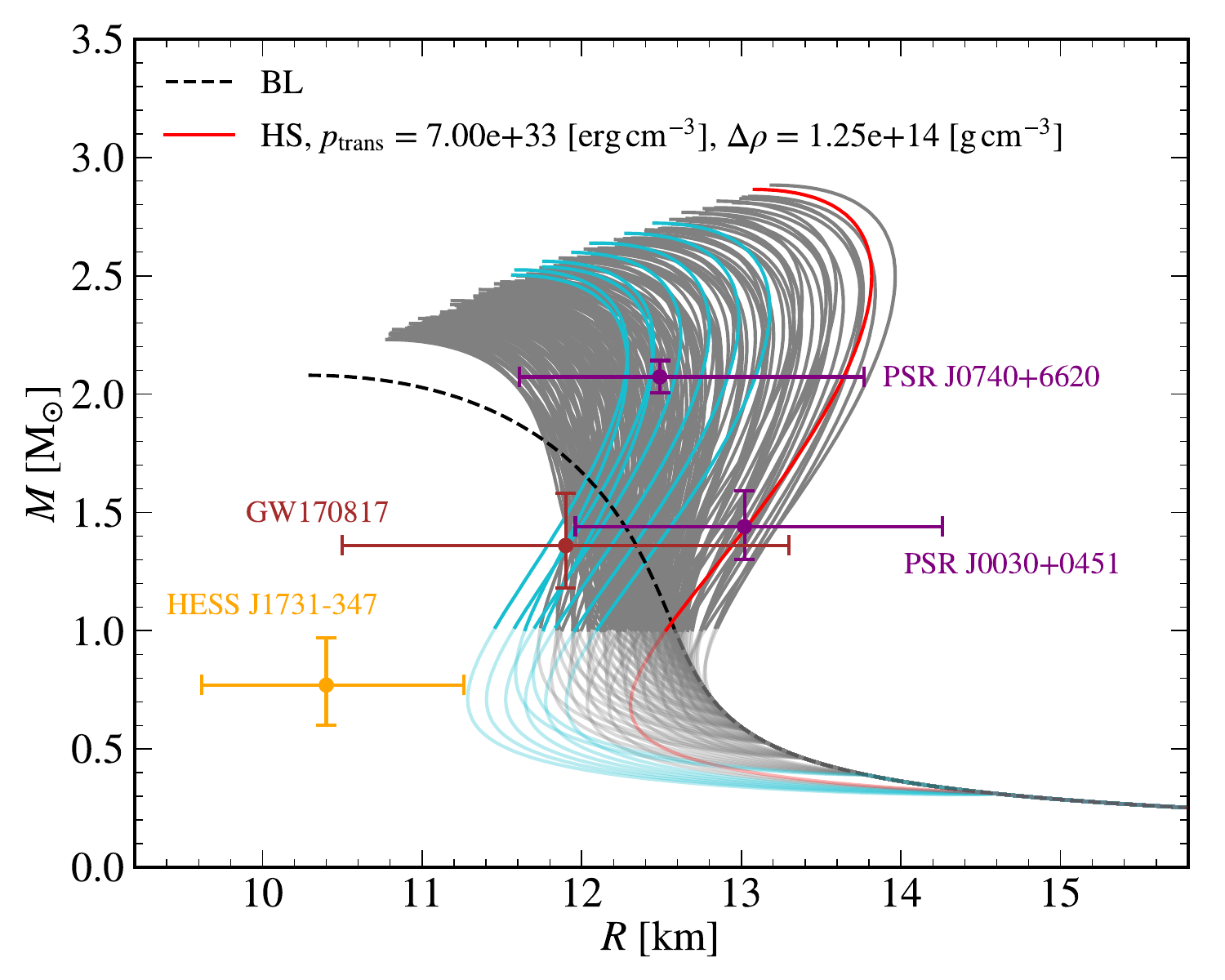}}
    \subfigure[$I$-$Q$ relations]{\includegraphics[width=0.95\columnwidth]{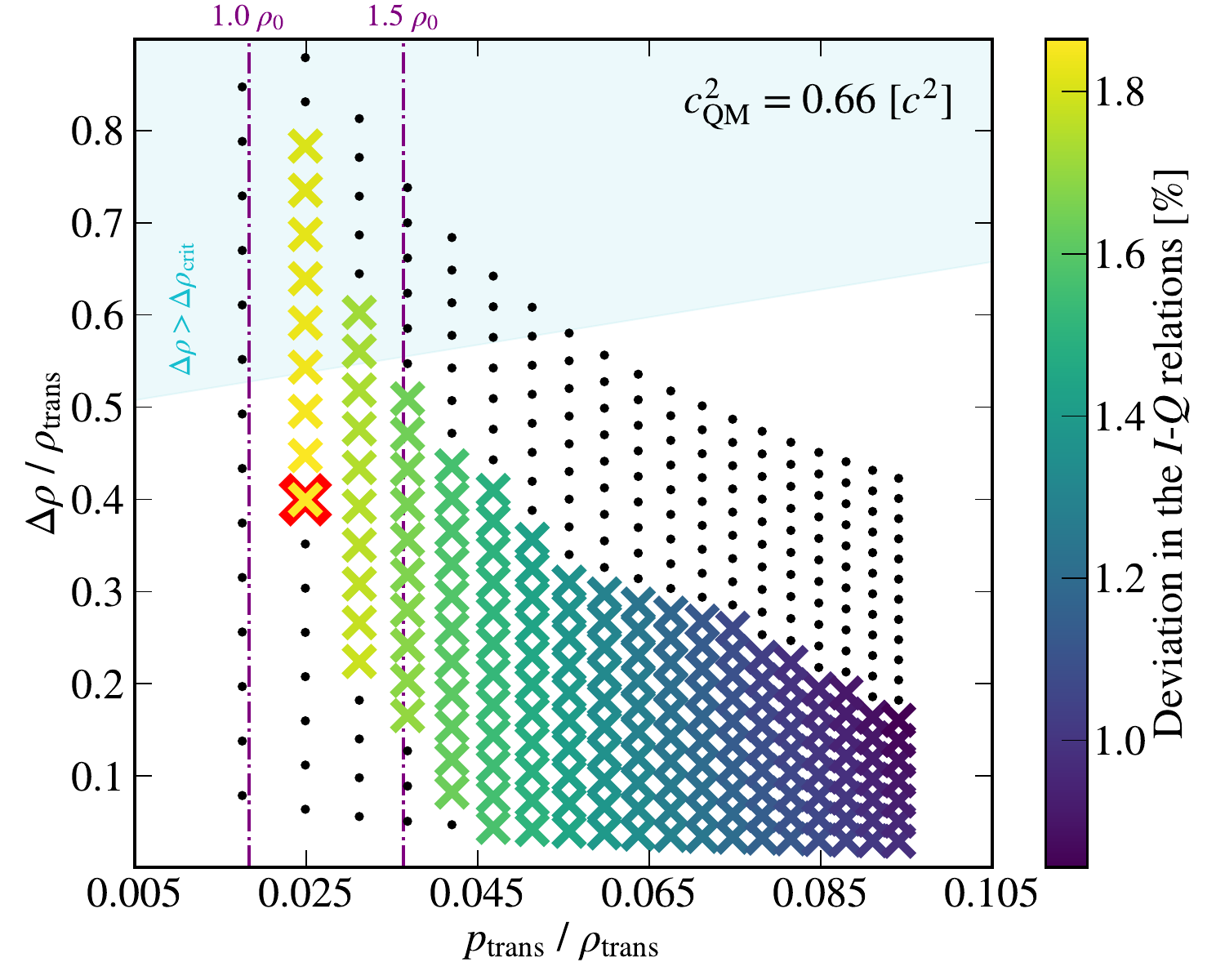}}
    \caption{
    Same as Fig.~\ref{fig: 2d, HS, APR, v=066}, but for HS models with the {BL} EOS being the NM part.
    }
    \label{fig: 2d, HS, BL, v=066}
\end{figure*}

\begin{figure*}[!ht]
    \subfigure[$M$-$R$ relations]{\includegraphics[width=0.95\columnwidth]{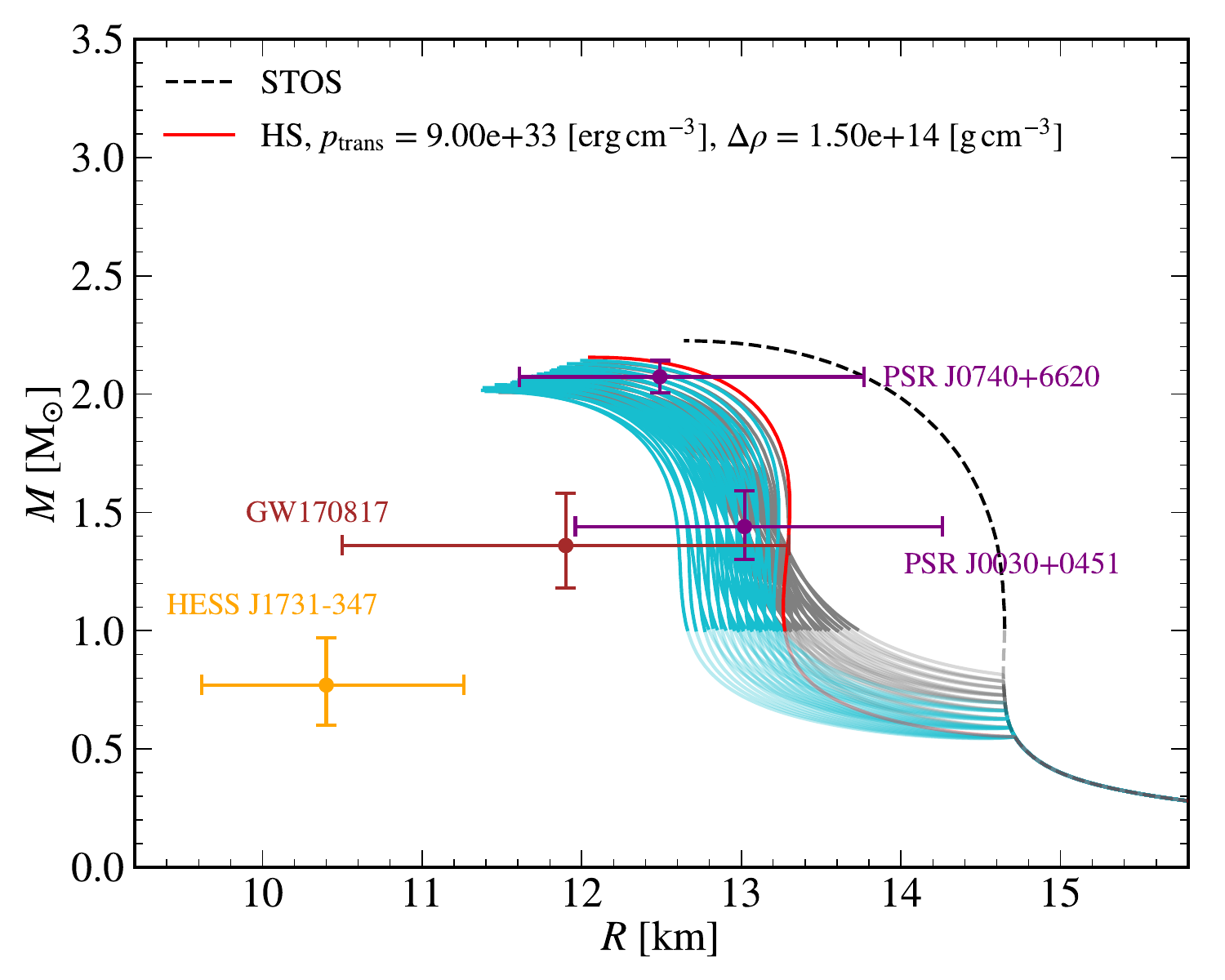}}
    \subfigure[$I$-$Q$ relations]{\includegraphics[width=0.95\columnwidth]{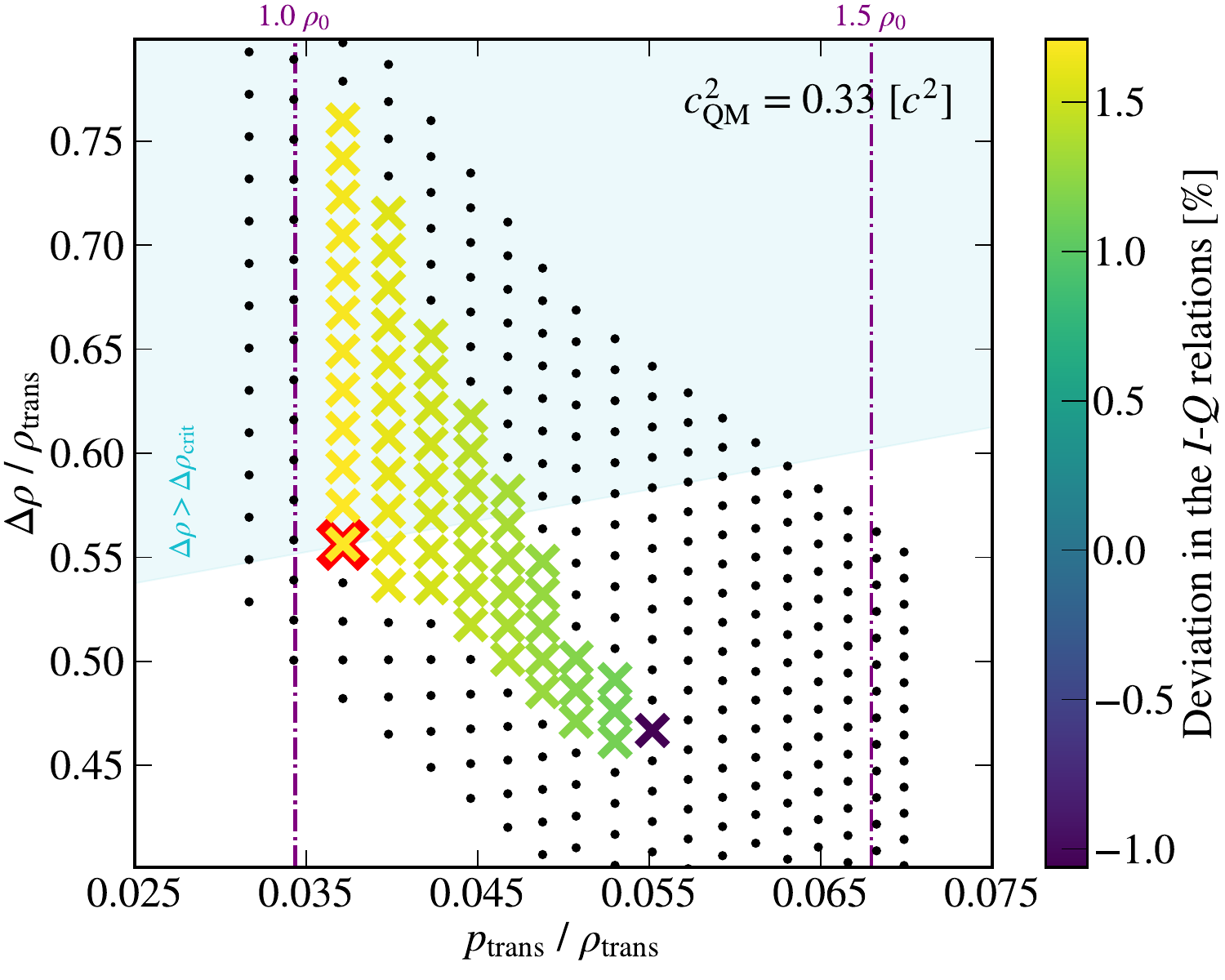}}
    \caption{
    Same as Fig.~\ref{fig: 2d, HS, APR, v=033}, but for HS models with the {STOS} EOS being the NM part.
    }
    \label{fig: 2d, HS, STOS, v=033}
\end{figure*}

\begin{figure*}[!ht]
    \subfigure[$M$-$R$ relations]{\includegraphics[width=0.95\columnwidth]{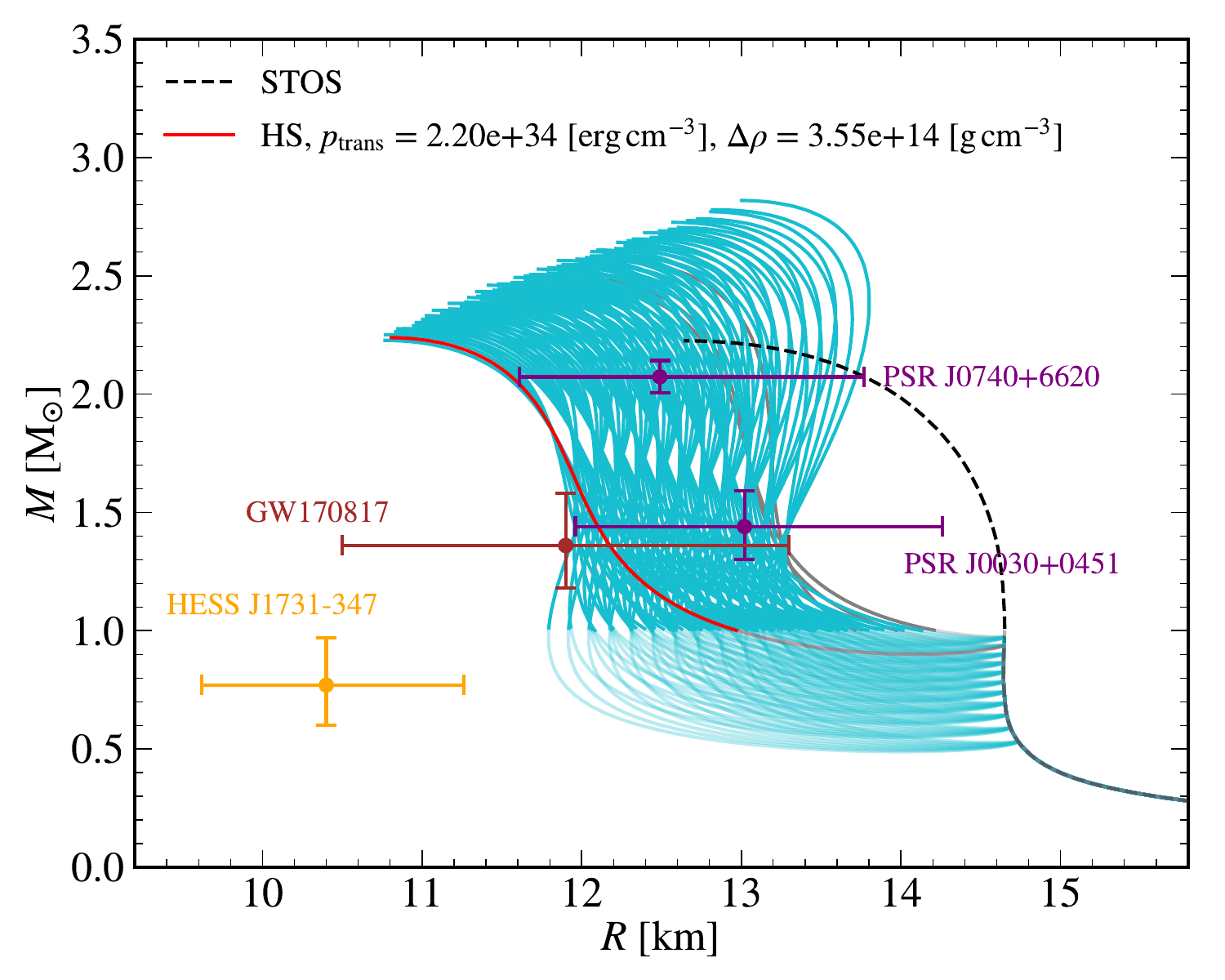}}
    \subfigure[$I$-$Q$ relations]{\includegraphics[width=0.95\columnwidth]{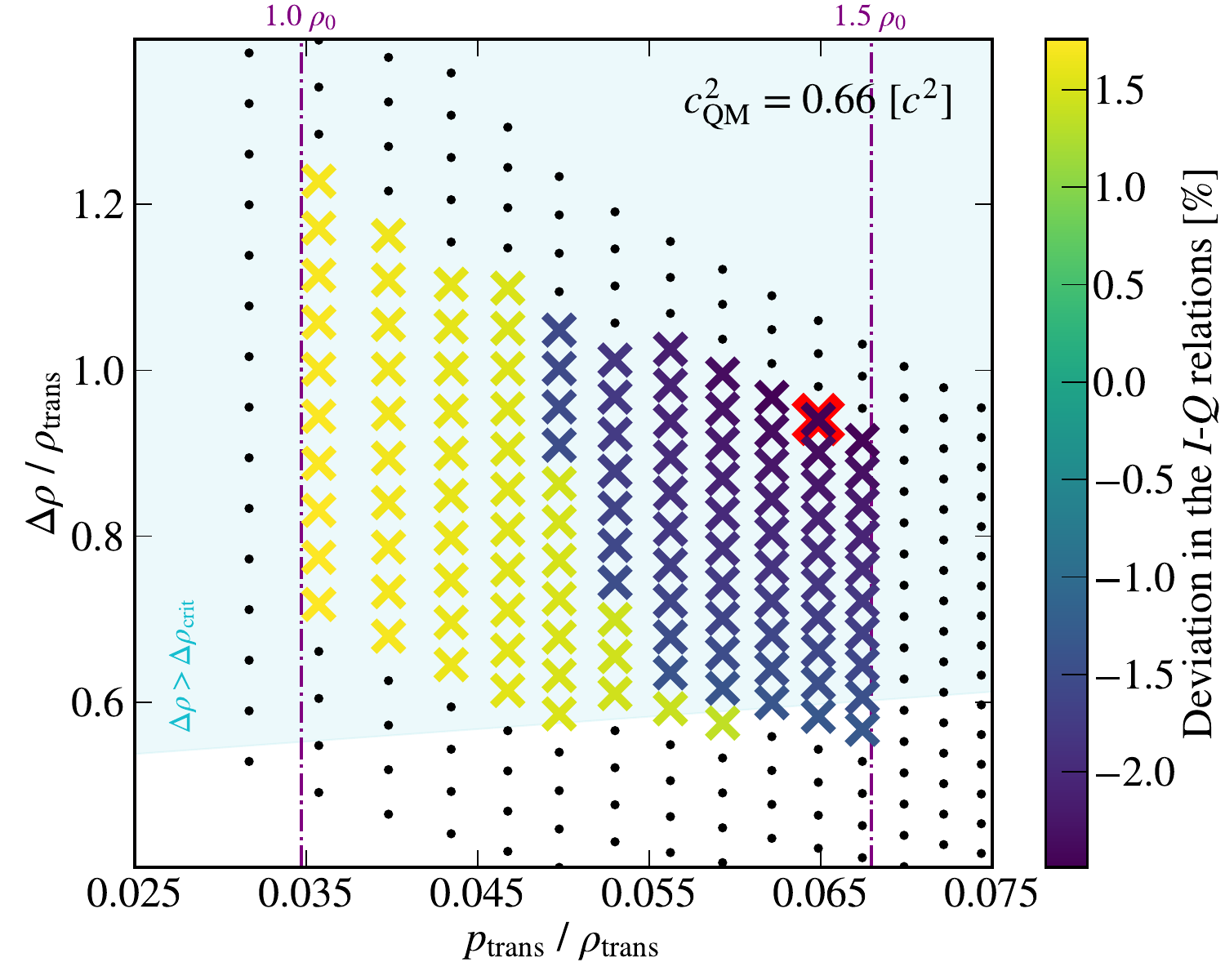}}
    \caption{
    Same as Fig.~\ref{fig: 2d, HS, APR, v=066}, but for HS models with the {STOS} EOS being the NM part.
    }
    \label{fig: 2d, HS, STOS, v=066}
\end{figure*}

The canonical HS model adopts the {APR} EOS as the NM part and $\{ p_\text{trans}, \Delta\rho, c_\text{QM}^2 \} = \{ 6.00\times10^{33}{\ecc}, 2.00\times10^{13}{\gcc}, 0.33{\cs} \}$ for the CSS parameters.
The corresponding transition density of this hybrid EOS on the NM side is $1.34${\rhoo}, and the TOV limit is $\approx2.2${\msol}, similar to that of a NS with the same NM EOS. 
In the following, we discuss only the behavior of configurations above $1${\msol} that are astrophysically relevant. 
The $M$-$R$ relations for the canonical HS models with $\kappa$ up to $7\times10^{26}${\pa} are shown in Fig.~\ref{fig: M-R, HS, APR, vary kappa}. 
The presence of the QM shear modulus effectively stiffens the EOS, and thus a more massive HS can be supported.
The TOV limit is enhanced by $\approx5\%$ when the maximum value of $\kappa$ is considered compared to the fluid limit, qualitatively and quantitatively in agreement with the results reported in Ref.~\cite{Dong:2024lte}.

Next, we analyze the effects of the QM shear modulus on the universal relations for the HS models.
Figure~\ref{fig: vs_pc-4, HS, APR, vary kappa} shows that the presence of the QM shear modulus always increases $C$, as well as decreases $\bar{I}$, $\bar{\lambda}_2$, and $\bar{Q}$ for an elastic star at any given central pressure $p_c$.
Such effects are more pronounced for high-mass configurations with larger $p_c$, where the elastic QM cores are more massive.

The $I$-${\lambda}_2$-$Q$-$C$ relations for the above models are further displayed in Fig.~\ref{fig: uni, HS, APR, vary kappa}.
At the fluid limits, the $I$-${\lambda}_2$-$Q$ relations for the {APR} EOS model and the canonical HS model are universal, i.e., deviate from most of the other EOS, represented by the fitting formulas~\cite{Yagi:2016bkt}, by less than $1\%$. 
When the nonvanishing QM shear modulus is introduced to the canonical HS model, the $I$-${\lambda}_2$-$Q$ relation curves shift systematically.
Qualitatively, the $I$-${\lambda}_2$ and $I$-$Q$ relation curves shift upward, while the $Q$-${\lambda}_2$ relation curve shifts downward, when $\kappa$ increases.
In particular, the magnitude of the percentage relative difference of the $I$-$Q$ and $Q$-${\lambda}_2$ relations from the fitting formulas reaches $\approx2\%$ when the maximum value of $\kappa$ is considered.
Similar systemic shifts are observed for the universal relations related to $C$ as well for the elastic HS models, where the $I$-$C$, $C$-${\lambda}_2$, and $Q$-$C$ relation curves all shift upward when $\kappa$ increases. 
However, the deviation in these universal relations for elastic HS models ($\approx5\%$ at $\kappa = 7\times10^{26}${\pa}) is not significant, since the {APR} EOS and other realistic EOS models can already deviate from the fitting formulas by a similar magnitude~\cite{Yagi:2016bkt}. 

\subsubsection{Parameter scan}

Subsequently, we vary the CSS parameters from those adopted in the canonical HS model.
The $I$-$Q$ and $Q$-${\lambda}_2$ relations show similar magnitude of deviation in general, which is larger than that of the $I$-${\lambda}_2$ relations.
Hence, we focus on the discussions on the $I$-$Q$ relations to quantify the effects of the QM shear modulus in the following. 
We also skip the detailed discussions on the universal relations related to $C$ because the deviation owing to the QM shear modulus cannot noticeably overwhelm the uncertainty among different NM EOS. 
Keeping $c_\text{QM}^2 = 0.33{\cs}$ and varying $p_\text{trans}$ ($\Delta\rho$) linearly between $2\times10^{33-34}{\ecc}$ ($2\times10^{13-14}{\gcc}$), we construct a collection of $361$ hybrid EOS in total. 
We also construct another collection of hybrid EOS with $c_\text{QM}^2 = 0.66{\cs}$, where the varying range of $p_\text{trans}$ and $\Delta\rho$ are slightly adjusted to produce more models that satisfy the constraints listed in Sec.~\ref{sec: methods}.
The $M$-$R$ relations and the deviation in the $I$-$Q$ relations for the corresponding HS models are shown in Figs.~\ref{fig: 2d, HS, APR, v=033} and \ref{fig: 2d, HS, APR, v=066}, where $\kappa=7\times10^{26}${\pa} is assumed. 
As a general trend, the deviation in the $I$-$Q$ relations is more pronounced for smaller values of $p_\text{trans}$, $\Delta\rho$, and $c_\text{QM}^2$. 
For a fixed value of $c_\text{QM}^2$, smaller values of $p_\text{trans}$ and $\Delta\rho$ usually correspond to effectively stiffer EOS. 
Hence, the elastic HS model with the largest magnitude of deviation in the $I$-$Q$ relations usually appears near the upper-right edge in the $M$-$R$ relations among all models.
To access the largest possible magnitude of deviation, we additionally construct several hybrid EOS with even smaller values of these CSS parameters. 
At $c_\text{QM}^2 = 0.24{\cs}$, we find a few HS models with deviation $\approx2.7\%$ in the $I$-$Q$ relations. 
The hybrid EOS are too soft to produce massive enough models that are compatible with the $M$-$R$ relation measurement of the pulsar {PSR J0740+6620} with further smaller values of $c_\text{QM}^2$. 
Thus, we conclude that the $I$-${\lambda}_2$-$Q$ relations for the elastic HS models are valid up to $3\%$.

We remark that a few HS models presented in Figs.~\ref{fig: 2d, HS, APR, v=033} and \ref{fig: 2d, HS, APR, v=066} are also compatible with the $M$-$R$ relation for the central compact object within the supernova remnant {HESS J1731-347}. 
Also, some HS models in these figures lie in the region where the Seidov stability condition is violated, i.e., with $\Delta\rho > \Delta\rho_\text{crit}$ [see Eq.~\eqref{eq: Seidov stability condition} for the definition of $\Delta\rho_\text{crit}$].
In these exceptional cases, we do not observe significantly larger deviation in the universal relations.


\begin{figure}
    \includegraphics[width=0.95\columnwidth]{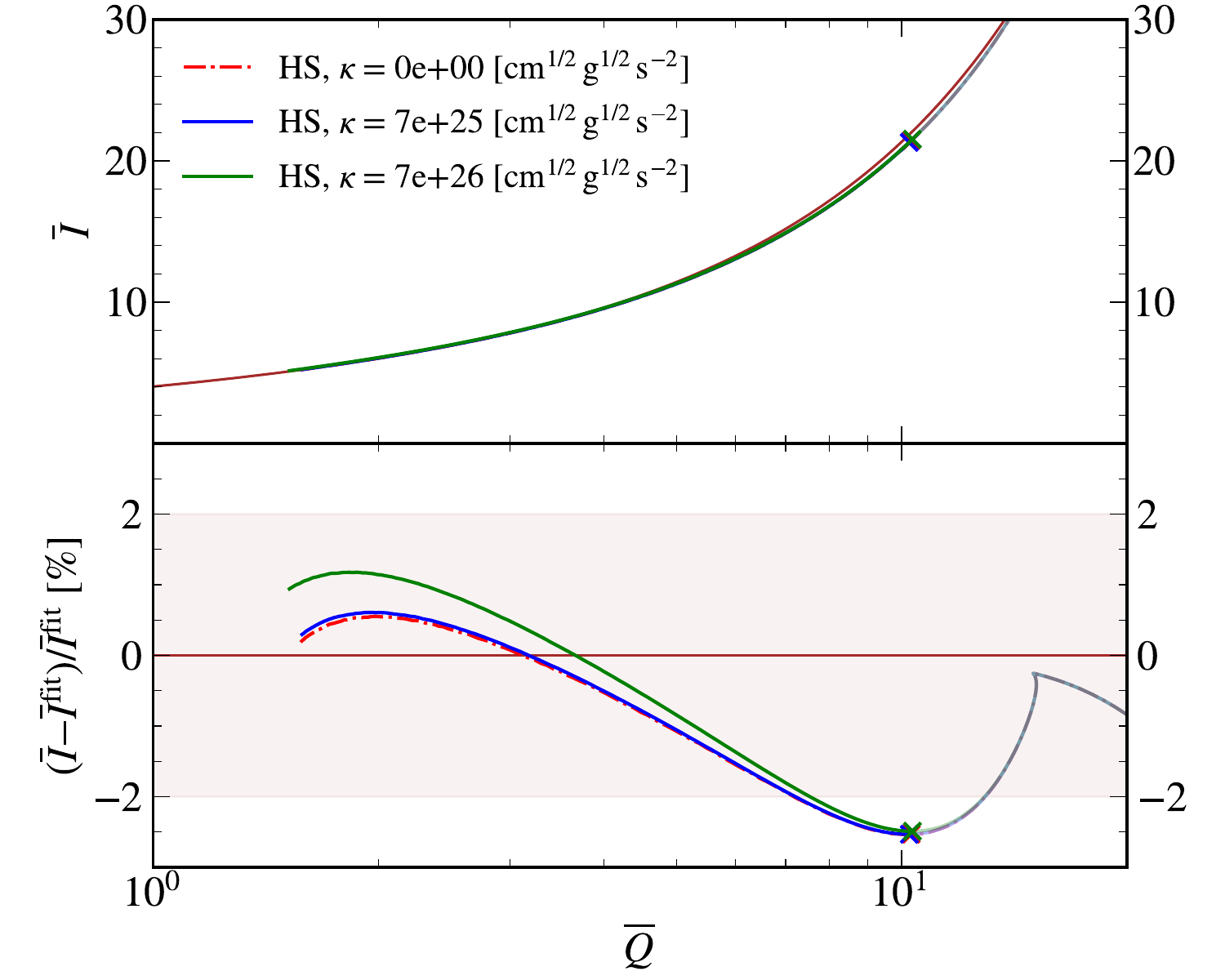}
    \caption{
    Same as in Fig.~\ref{fig: uni, HS, APR, vary kappa, I-Q}, but for the HS model highlighted in red displayed in Fig.~\ref{fig: 2d, HS, STOS, v=066} with different values of $\kappa$. 
    This model features a strong phase transition with $\Delta\rho$ considerably larger than $\Delta\rho_\text{crit}$. 
    A kink appears in the $I$-$Q$ relation curves at $\bar{Q} \approx 15$ where the phase transition occurs. 
    The deviation in the $I$-$Q$ relations reach the extrema at $\bar{Q} \approx 10$ near the phase transition point instead of the TOV limits.
    }
    \label{fig: uni, HS, STOS, vary kappa}
\end{figure}



To estimate the sensitivity of the universal relations for elastic HSs to the NM part of hybrid EOS, we repeat the above calculations using the {BL} and {STOS} EOS for constructing the hybrid EOS.
The {APR}, {BL}, and {STOS} EOS have low, intermediate, and high stiffness in terms of sound speed squared $c_s^2\equiv \dv{p}{\rho}$ below $\approx 3\,\rho_0$, respectively\footnote{Note that, however, the stiffness of the {APR} EOS surpasses the others above $\approx 3\,\rho_0$ and becomes superluminal before the TOV limit is reached.}.
The calculation results are shown in Figs.~\ref{fig: 2d, HS, BL, v=033}--\ref{fig: 2d, HS, STOS, v=066} for these new HS models. 

As a whole, the effects of the QM shear modulus on the new HS models using the {BL} EOS, such as the enhancement of the TOV limits and the magnitude of deviation in the universal relations, are similar to those on the original HS models using the {APR} EOS. 
In Figs.~\ref{fig: 2d, HS, STOS, v=033} and \ref{fig: 2d, HS, STOS, v=066}, meanwhile, we find that several HS models using the {STOS} EOS feature a strong phase transition, and hence a disconnected branch emerges in the $M$-$R$ relations. 
The deviation in the $I$-$Q$ relations change sign when the phase transition is sufficiently strong.
Figure~\ref{fig: uni, HS, STOS, vary kappa} shows the $I$-$Q$ relations for the HS model with $\{ p_\text{trans}, \Delta\rho, c_\text{QM}^2 \} = \{ 2.20\times10^{34}{\ecc}, 3.55\times10^{13}{\gcc}, 0.66{\cs} \}$, i.e., the particular model with the largest magnitude of deviation in the $I$-$Q$ relations, as an example to elucidate such a phenomenon. 
At the onset of phase transition, a kink appears in the $I$-$Q$ relation curve at $\bar{Q} \approx 15$, and the curve bends downward drastically until $\bar{Q} \approx 10$.
Even at the fluid limit, the magnitude of deviation reaches $\approx2.5\%$ where the bending curve reaches the minimum.
In fact, it has been shown that fluid HSs with a strong phase transition have an uncertainty up to $\sim3\%$ in the $I$-${\lambda}_2$-$Q$ relations~\cite{Paschalidis:2017qmb}, which supports our finding in Fig.~\ref{fig: uni, HS, STOS, vary kappa}.
On the other hand, the general effects of the QM shear modulus, as illustrated in Fig.~\ref{fig: uni, HS, APR, vary kappa, I-Q}, are to shift the $I$-$Q$ relation curves upward. 
For this particular model, the deviation in the $I$-$Q$ relations reach the extremum near the phase transition point, in contrast to many other HS models aforementioned that reach the extrema near the TOV limits, even if $\kappa = 7\times10^{26}${\pa} is considered. 
The extremum near the phase transition point is not related to the QM shear modulus, which mainly affects the $I$-$Q$ relations for the high-mass configurations.
For the majority of elastic HS models without featuring a strong phase transition, the universal relation curves are qualitatively similar to those of the canonical HS model presented in Fig.~\ref{fig: uni, HS, APR, vary kappa}.
In summary, we believe our conclusions on the general impacts of the QM shear modulus on the elastic stars do not depend sensitively on the NM part of the hybrid EOS.

\subsection{Quark stars}

\begin{figure*}
    \subfigure[$M$-$R$ relations]{\includegraphics[width=0.95\columnwidth]{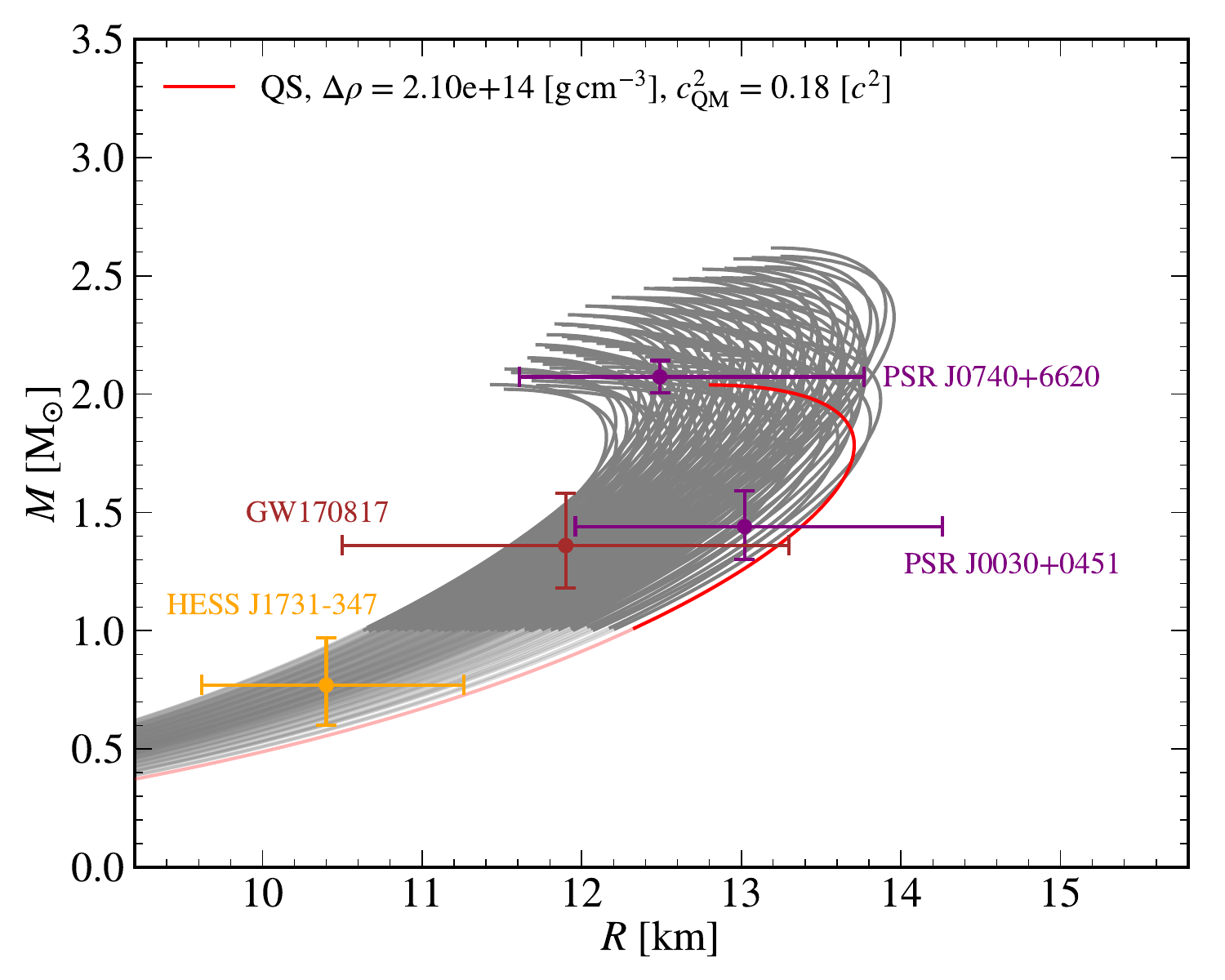}}
    \subfigure[$I$-$Q$ relations]{\includegraphics[width=0.95\columnwidth]{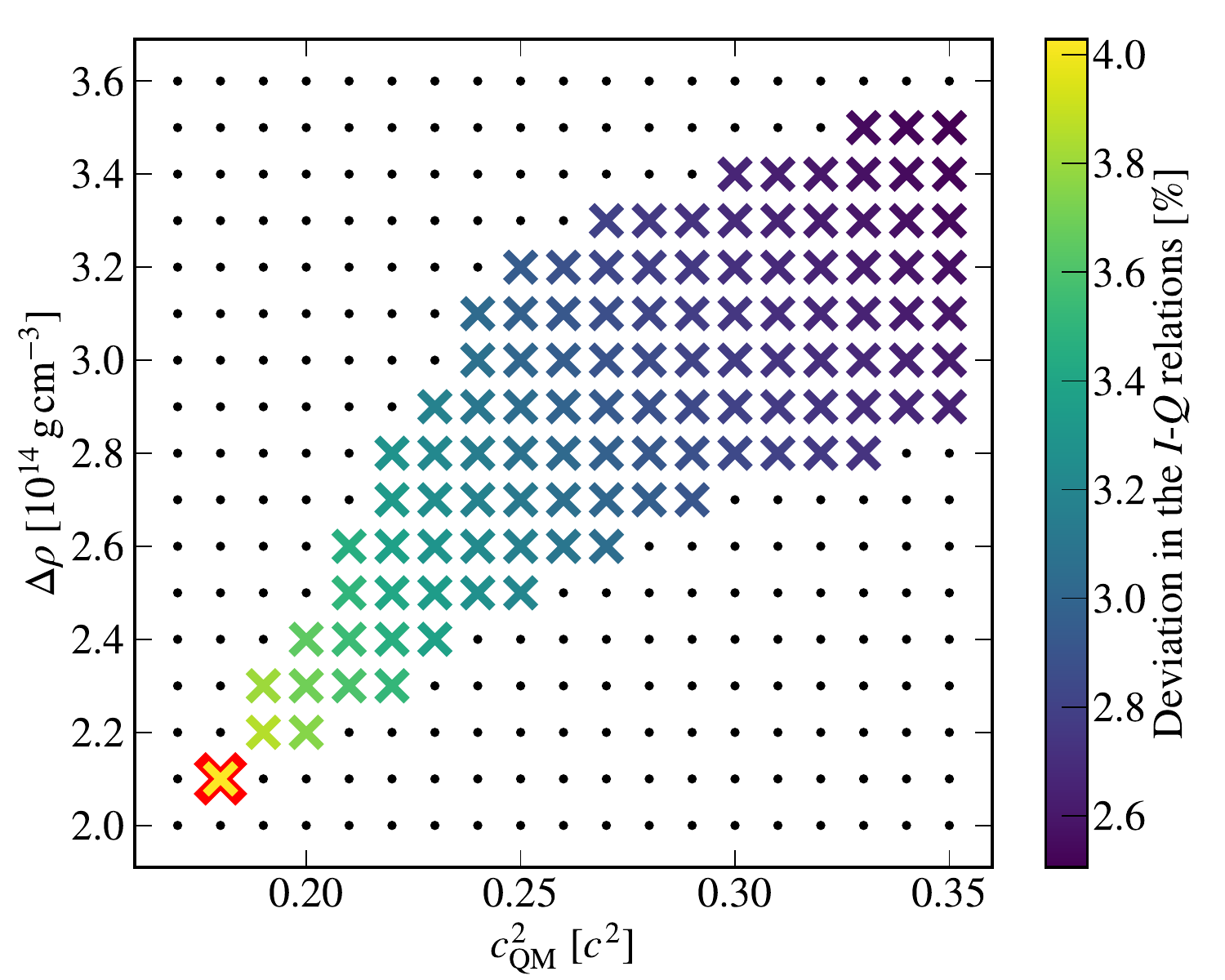}}
    \caption{
    (a) Same as in Fig.~\ref{fig: 2d, HS, APR, v=033, M-R}, but for QS models with varying values of $c_\text{QM}^2$ and $\Delta\rho$.
    (b) Same as in Fig.~\ref{fig: 2d, HS, APR, v=033, I-Q}, but in $c_\text{QM}^2$-$\Delta\rho$ plane.
    }
    \label{fig: 2d, QS}
\end{figure*}

\begin{figure}
    {\includegraphics[width=0.95\columnwidth]{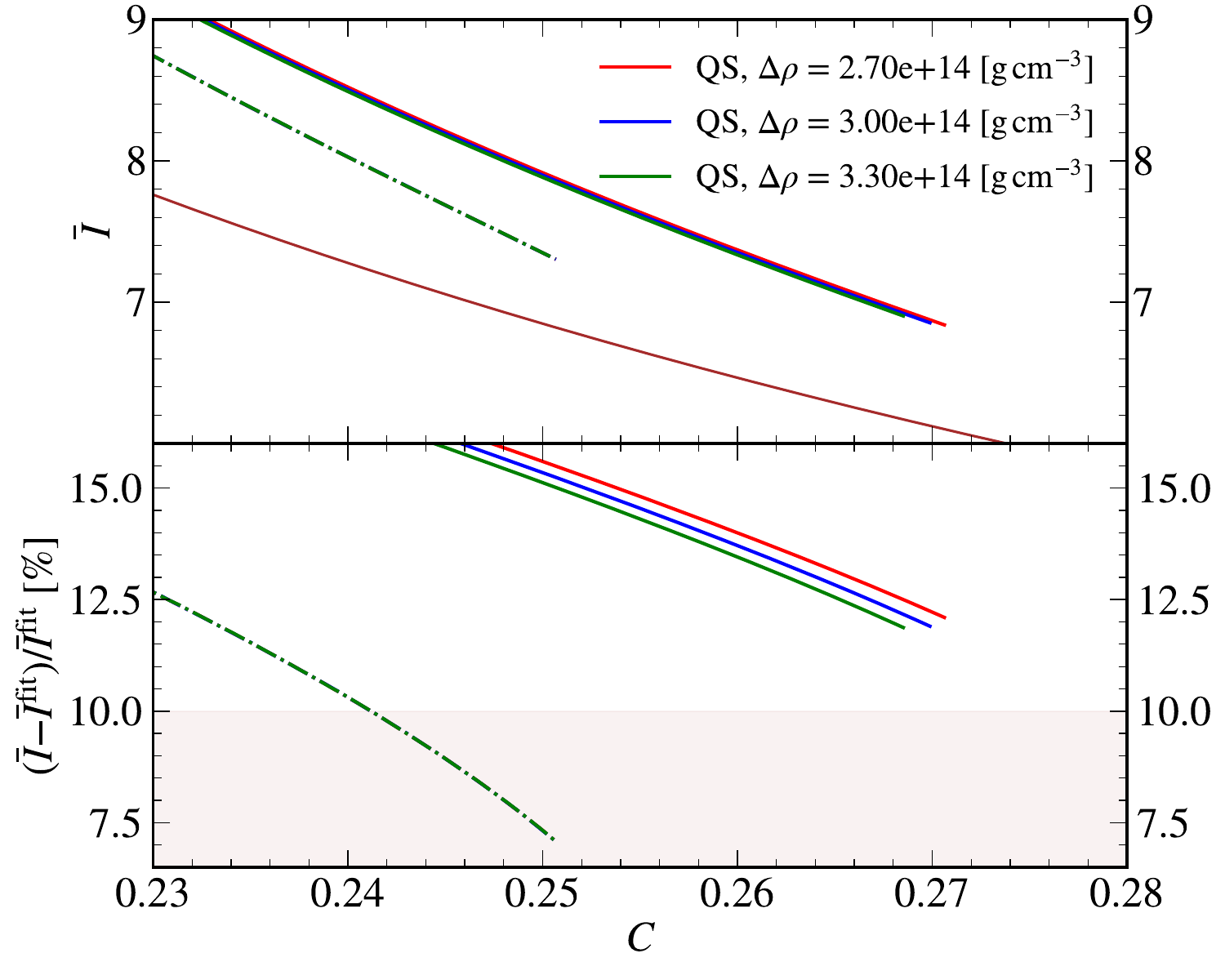}}
    \caption{
    Same as in Fig.~\ref{fig: uni, HS, APR, vary kappa, I-C}, but for a few QS models with $c_\text{QM}^2 = 0.27{\cs}$ and different values of $\Delta\rho$. 
    The dash-dotted curve (overlapping with each other in the figure) represents the fluid limits, and the solid curves represent the elastic models with $\kappa = 7\times10^{26}${\pa}.
    This figure focuses on the curves near the TOV limits, where the variation among the solid curves becomes noticeable.
    }
    \label{fig: uni, QS, I-C}
\end{figure}

We next study QS models. 
We set $p_\text{trans} = 0$ and vary the other two CSS parameters\footnote{Such a choice results in QS models with a surface density equals $\Delta\rho$.}. 
The $M$-$R$ and $I$-$Q$ relations for such a collection of elastic QS models with $\kappa = 7\times10^{26}${\pa} are shown in Fig.~\ref{fig: 2d, QS}.
In comparison to the previous section regarding the universal relations for the elastic HSs, we observe slightly larger deviation in these relations for the elastic QSs. 
Although these QS models can satisfy the $I$-${\lambda}_2$-$Q$ universality at the fluid limits, the deviation can reach $\approx4\%$ when $\kappa=7\times10^{26}${\pa} is considered. 
Similar to the canonical HS model, the $I$-${\lambda}_2$ and $I$-$Q$ relation curves for the QS models shift upward, and the $Q$-${\lambda}_2$ relation curve shifts downward when $\kappa$ increases.
The above comparison indicates that the systematic shifts in the $I$-${\lambda}_2$-$Q$ relation curves do not qualitatively depend on the presence of the fluid NM envelope. 

The universal relations related to $C$ are less robust for low-mass QSs, which have significantly smaller radii, and hence larger $C$, relative to the radius of NSs at the same mass.
Within the CSS parametrization framework, QS models with different surface densities can be constructed by tuning the value of $\Delta\rho$ while keeping $c_\text{QM}^2$ constant. 
Figure~\ref{fig: uni, QS, I-C} shows the $I$-$C$ relations for a few QS models as an example.  
They adopt different values of $\Delta\rho$ but the same value of $c_\text{QM}^2 = 0.27{\cs}$.
From our numerical results, the $I$-$C$ relations for these QS models are almost identical at the fluid limits (see Refs.~\cite{Chan:2014tva,Chan:2015iou} for relevant discussions from the post-Minkowskian expansion approach and the connection to incompressible stars). 
For the same set of QS models with $\kappa=7\times10^{26}${\pa}, on the other hand, the curves vary by $\sim1\%$ near the TOV limits, i.e., the presence of the QM shear modulus breaks the degeneracy of these QS models with different values of $\Delta\rho$ but same $c_\text{QM}^2$, introducing systematic shifts to the curves with respect to the fluid limits.
Likewise, variations in the $C$-${\lambda}_2$ and $Q$-$C$ relation curves with a similar magnitude are observed for these elastic QS models.

Finally, we remark that the magnitude of deviation in the universal relations for elastic stars reported above is obtained based on the calculation results of the QM shear modulus by Ref.~\cite{Mannarelli:2007bs}. 
If an even larger value of QM shear modulus is realized in future studies, one would expect even larger deviation in these universal relations for elastic stars in our formulation.

\section{Conclusions and Discussions}
\label{sec: conclusions and discussions}

In this paper, we investigated the $I$-${\lambda}_2$-$Q$-$C$ relations for elastic HS and QSs in the slow rotation and small tidal deformation approximation. 
The corresponding background and perturbation equations were obtained using the stress-energy tensor with scalar pressure anisotropy.
In comparison to the earlier study of Ref.~\cite{Yagi:2015hda} employing the same formulation, where phenomenological anisotropy models were assumed, we adopted a fully relativistic theory of elasticity to describe the nonlinear stress-strain relation~\cite{Karlovini:2002fc}. 
Hence, we were able to construct the sheared background of elastic stars in a self-consistent manner, with a physical origin of pressure anisotropy associated with the QM in the CCS phase.
As a whole, we found that the deviation in the $I$-${\lambda}_2$-$Q$-$C$ relations for our elastic star models from the fluid limits is smaller than the results reported in Ref.~\cite{Yagi:2015hda} within the anisotropy range of $\lambda_\mathrm{H}$, $\lambda_\mathrm{BL} = [-2,2]$. 

The studies in Refs.~\cite{Lau:2017qtz,Lau:2018mae,Dong:2024opz} investigated the $I-\lambda_2$ and $C-\lambda_2$ relations for elastic HS and QSs containing the CCS QM with unsheared background configuration models. 
Although $\lambda_2$ of these elastic stars were significantly reduced relative to the fluid limits, $I$ and $C$ remained unchanged. 
Thus, a large deviation was found in their $I-\lambda_2$ and $C-\lambda_2$ relations from the fluid limits up to tens of percentages, purely due to the reduction in $\lambda_2$.
For our elastic star models with a sheared background, on the other hand, all these quantities undergo systematic shifts relative to the fluid limits at any $p_c$ (see Fig.~\ref{fig: vs_pc-4, HS, APR, vary kappa}). 
Take the $C-\lambda_2$ relation as an example. 
In Ref.~\cite{Dong:2024opz}, Fig.~1 shows the deviation in the $C-\lambda_2$ relations for their elastic HS models from the fitting formula for fluid NSs (the same fitting formula as we used in this paper). 
The curves representing the elastic HS models lie below the fitting formula because $\lambda_2$ is remarkably reduced by the QM shear modulus, whereas $C$ remains compatible with typical fluid NS models. 
In Fig.~\ref{fig: uni, HS, APR, vary kappa}, in contrast, the $C-\lambda_2$ relation curves for our elastic HS models are above the fitting formula as well as the fluid limit curve, in which $\lambda_2\,(C)$ decreases (increases) when the effects of the QM shear modulus are taken into account. 
The resulting deviation in the $C-\lambda_2$ relations from the fitting formula for our models, therefore, is qualitatively different from that reported in Ref.~\cite{Dong:2024opz}. 
Also, the magnitude of deviation in our cases is somehow canceled out by the mutual shift in $C$ and $\lambda_2$ of the elastic stars. 


The results in this paper appear to be different from the conclusion of previous work on solid QSs, where the $I-\lambda_2$ relations are found to significantly deviate from the fluid case by up to $\sim 30\%$. 
This deviation can either arise from the assumption in the previous work, where the background star is taken to be unsheared, or from the assumption in the formalism we employ here.
Strictly speaking, the formulation we employ here for the nonradial perturbations (i.e., on determining $\lambda_2$ and $Q$) does not fully capture the elastic deformation. 
In particular, some off-diagonal terms in the spatial part of the perturbed stress-energy tensor are missing due to a restrictive assumption on the solid EOS in the perturbed configuration\footnote{The stress-strain relation should depend on the eigenvalues of the strain tensor in the perturbed star under the eigenbasis formulation by Ref.~\cite{Karlovini:2007}, resulting in $(r,\theta)$, $(r,\phi)$, $(\theta,\phi)$ components in the stress-energy tensor, which are not present if we assume that the perturbed radial vector $k^\alpha$ remains orthogonal to the surface of the two-sphere as in Ref.~\cite{Yagi:2015hda}.}. 
However, these missing components are proportional to the transverse sound speeds (see Ref.~\cite{Karlovini:2007}), which are much smaller than the perturbations in the diagonal components that scale as the longitudinal sound speeds~\cite{Dong:2024lte}, even at the same perturbative level. 
Therefore, we expect the results reported here to serve as a reasonable first approximation of the true elastic deformations based on the existing formalism for anisotropic stars. 
Our next goal is to develop the full perturbation equations for polar deformations based on relativistic elastic theory~\cite{Carter:1973, Karlovini:2007}.

Another important direction for future work is to investigate the origin of universality. 
Reference~\cite{Yagi:2014qua} pointed out that isodensity contours inside stars become increasingly self-similar as $C$ grows. 
This observation suggests that an approximate, emergent symmetry may underlie the origin of universality. 
The authors further showed that the EOS variation in the $I$-${\lambda}_2$-$Q$ relations increases as the isodensity contours deviate from self-similarity. 
Since self-similarity becomes exact in the limit of incompressible stars, this evidence supports the idea that universality arises because realistic NSs are effectively "close to" incompressible configurations. 
This conjecture was further strengthened in Ref.~\cite{Chan:2014kua}, where the authors explicitly demonstrated that the EOS dependence in the universal relations between ${\lambda}_2$ and the fundamental-mode oscillation frequency is suppressed by adopting a modified Tolman VII model in the Newtonian limit and perturbing about the incompressible limit (see also Refs.~\cite{Sham:2014kea,Jiang:2020uvb,Kyutoku:2025zud,Katagiri:2025qze} for additional studies exploring possible origins of universality in various relations). 
It would be interesting to determine whether the origin of universality can be understood at a more fundamental level.


\begin{acknowledgments}
We thank Zoey Zhiyuan Dong for helpful discussions. 
C. M. Y. is supported by grants from the Research Grant Council of the Hong Kong
Special Administrative Region, China (Projects No. 14300320 and No. 14304322) and the European Union
through ERC Synergy Grant HeavyMetal No. 101071865. 
S. Y. L. acknowledges
support from Montana NASA EPSCoR Research Infrastructure Development under Award No. 80NSSC22M0042.
K. Y. acknowledges support from NSF Grant No. PHY2339969 and the Owens Family Foundation.
\end{acknowledgments}

\appendix

\section{Background and Perturbation Equations}
\label{app: background and perturbation equations}

In this appendix, we review how to obtain the perturbation equations of a slowly rotating, tidally deformed elastic star with a static, spherically symmetric, and sheared background.
We introduce metric perturbations $\delta\omega$, $\delta h$, $\delta k$, and $\delta m$, up to the second order in slow rotation, to the metric ansatz of a static, spherically symmetric star in the Schwarzschild coordinates $(t,\Bar{r},\theta,\phi)$~\cite{Yagi:2013awa,Yagi:2015hda}, 
\begin{equation}
\begin{aligned}
    \dd s^2 &= - e^\nu \left( 1 + 2 \epsilon^2 \delta h \right) \dd t^2 + e^{\lambda} \left( 1 + \frac{2 \epsilon^2 \delta m}{\Bar{r} - 2 m} \right) \dd \Bar{r}^2 \\
    &+ \Bar{r}^2 \left( 1 + 2 \epsilon^2 \delta k \right) \bigl\{ \dd \theta^2 + \sin^2\theta \left( \dd \phi - \epsilon \left( \Omega - \delta\omega \right) \right]^2 \bigl\} \\
    &+ \mathcal{O}\left(\epsilon^3\right),
    \label{eq: metric element}
\end{aligned}
\end{equation}
where $\epsilon$ is a bookkeeping parameter to count the order of slow rotation and $\Omega$ is the angular frequency of the star. 
$\nu(\Bar{r})$ and $\lambda(\Bar{r})$ are the metric functions, where $e^{\lambda}$ can be written as
\begin{equation}
    e^{\lambda(\Bar{r})} = \left[ 1 - \frac{2 m(\Bar{r})}{\Bar{r}} \right]^{-1},
\end{equation}
with $m$ being the gravitational mass enclosed within a sphere of radius $\Bar{r}$. 

We decompose the metric perturbations in terms of Legendre polynomials, 
\begin{align}
    \delta \omega(\Bar{r}, \theta) &= \omega_1 (\Bar{r}) P_{1}^{\prime}(\cos \theta), \\
    \delta h(\Bar{r}, \theta) &= h_0 (\Bar{r}) + h_2 (\Bar{r}) P_{2}(\cos \theta), \\
    \delta m(\Bar{r}, \theta) &= m_0 (\Bar{r}) + m_2 (\Bar{r}) P_{2}(\cos \theta), \\
    \delta k(\Bar{r}, \theta) &= k_2 (\Bar{r}) P_{2}(\cos \theta),
\end{align}
where $P_l (\cos \theta)$ is the $l$th order Legendre polynomial and $P_{l}^{\prime}(\cos \theta) = \dd P_{l}(\cos \theta) / \dd \cos \theta$. 
Further, we introduce a new radial coordinate $r$ via a new function $\xi$, 
\begin{equation}
    \Bar{r}(r, \theta) = r + \epsilon^2 \xi(r, \theta) + \mathcal{O}\left(\epsilon^3\right),
\end{equation}
and the Legendre decomposition of $\xi$ is given by
\begin{equation}
    \xi(r, \theta) = \xi_0 (r) + \xi_2(r) P_{2}(\cos \theta).
\end{equation}
We introduce the above radial coordinate transformation such that the perturbed energy density $\rho[\Bar{r}(r, \theta)]$ is the same as the unperturbed energy density at $r$.
By substituting Eq.~\eqref{eq: metric element} and the stress-energy tensor given by Eq.~\eqref{eq: stress-energy tensor} into the Einstein equations, we obtain the background and perturbation equations for the elastic star from the equation of motion. 

\subsection{Background equations}

At $\mathcal{O}\left(\epsilon^0\right)$, we obtain the TOV equations from the equation of motion, 
\begin{align}
    \dv{m}{r} &= 4 \pi r^2 \rho, \\
    \dv{\nu}{r} &= 2 \frac{4 \pi r^3 p_r + m}{r \left( r - 2 m \right)}, \\
    \dv{p_r}{r} &= -\left( \rho + p_r \right) \frac{4 \pi p_r r^3 + m}{r \left( r - 2 m \right)} - \frac{2 \sigma_0}{r}.
\end{align}
Here, we introduce the background anisotropy $\sigma_0 \equiv p_r - p_t$ to denote the difference between $p_r$ and $p_t$. 
Following Refs.~\cite{Karlovini:2002fc,Dong:2024lte}, the TOV equations are further reformulated into a new set of differential equations, with $\{ m, \tilde{p}, z \}$ being independent variables, 
\begin{align}
    \dv{m}{r} &= 4 \pi r^2 \rho, \label{eq: TOV(reformulated) m}\\
    \dv{\tilde{p}}{r} &= \frac{\tilde{\beta}}{r \beta_r} \biggl\{ - \left( \rho + p_r \right) \frac{4 \pi p_r r^3 + m}{r - 2 m} - 2 \sigma_0 \nonumber\\
    &+ 4 \left( e^{\lambda/2} z - 1 \right) \left[ \tilde{\mu} + 3 \tilde{\mu} S^2 + \frac{\sigma_0}{2} \left( \tilde{\Omega} - 1 \right) \right] \biggl\}, \label{eq: TOV(reformulated) p}\\
    \dv{z}{r} &= \frac{z}{r} \left[ \frac{r}{\tilde{\beta}} \dv{\tilde{p}}{r} - 3 \left( e^{\lambda/2} z - 1 \right) \right], \label{eq: TOV(reformulated) z}
\end{align}
where
\begin{align}
    \sigma_0 &= - \frac{\tilde{\mu}}{2} \left( z^{-2} - z^2 \right), \label{eq: sigma0}\\
    S^2 &= \frac{1}{6} \left( z^{-1} - z \right)^2, \label{eq: S2}\\
    p_r &= \tilde{p} + \left( \tilde{\Omega} - 1 \right) \tilde{\mu}S^2 + \frac{2}{3}\sigma_0, \label{eq: pr}\\
    \tilde{\beta} &= \left( \tilde{\rho} + \tilde{p} \right) \dv{\tilde{p}}{\tilde{\rho}}, \\
    \beta_r &= \tilde{\beta} + \frac{4}{3} \tilde{\mu} + \left[ \tilde{\Omega} \left( \tilde{\Omega} - 1 \right) + \tilde{\beta} \dv{\tilde{\Omega}}{\tilde{p}} \right] \tilde{\mu}S^2 \nonumber\\
    &+ 4 \left[ \tilde{\mu}S^2 + \frac{\sigma_0}{3} \left( \tilde{\Omega} - \frac{1}{2} \right) \right], \\
    \tilde{\Omega} &= \frac{\tilde{\beta}}{\tilde{\mu}} \dv{\tilde{\mu}}{\tilde{p}}.
\end{align}
Here, $z = n_r / n_t$, where $n_r\,(n_t)$ is the linear particle number density in radial (tangential) direction.
The variables with an overhead tilde represent the unsheared components.

We integrate Eqs.~\eqref{eq: TOV(reformulated) m}--(\ref{eq: TOV(reformulated) z}) under the following boundary conditions to obtain the background and $\sigma_0$ profiles of the elastic star. 
The boundary conditions near $r=0$ are given by
\begin{align}
    m &= \frac{4}{3} \pi \rho_c r^3 + \mathcal{O}(r^4), \\
    \tilde{p} &= p_c - 2 \pi \tilde{\beta}_c \frac{\left( \rho_c + p_c \right) \left( \rho_c + 3 p_c \right) - 4 \tilde{\mu}_c \rho_c}{3 \left( \tilde{\beta}_c + \frac{4}{3} \tilde{\mu}_c \right)} r^2 \nonumber\\
    &+ \mathcal{O}\left(r^3\right), \\
    \tilde{z} &= 1 - 4 \pi \frac{\left( \rho_c + p_c \right) \left( \rho_c + 3 p_c \right) + 3 \tilde{\beta}_c \rho_c}{15 \left( \tilde{\beta}_c + \frac{4}{3} \tilde{\mu}_c \right)} r^2 \nonumber\\
    &+ \mathcal{O}\left(r^3\right),
\end{align}
where the variables with a subscript $c$ represent the values at the center, which are unsheared by construction. 
From the solid quark core to the fluid envelope in the HS models (or the vacuum in the QS models), $\sigma_0$ is discontinuous. 
By requiring $p_r$ to be continuous across the core-envelope interface, we substitute Eqs.~\eqref{eq: sigma0} and~\eqref{eq: S2} into Eq.~\eqref{eq: pr} to obtain the junction condition on $\tilde{p}$, 
\begin{equation}
    \tilde{p}_{+} = \tilde{p}_{-} - \frac{\tilde{\mu}_{-}}{6} \left[ \left( 1 - \tilde{\Omega}_{-} \right) \left( z^{-1}_{-} - z_{-} \right)^2 + 2 \left( z^{-2}_{-} - z^2_{-} \right) \right],
\end{equation}
where the variables with a subscript $-$ $(+)$ represent the values evaluated at the core (envelope) side of the interface. 
For the HS models with a fluid envelope, $\tilde{p}_{+} = p_\text{trans}$ at the outer side of the interface; for the QS models, $\tilde{p}_{+} = 0$ as the vacuum is reached directly outside the solid quark core.
We refer the readers to Refs.~\cite{Karlovini:2002fc,Dong:2024lte} for more discussions on this junction condition. 
The boundary condition on $\tilde{p}$ at the stellar surface is given by 
\begin{equation}
    p_r = 0, \\
\end{equation}
i.e., a vanishing radial pressure. 
The total gravitational mass $M$ and radius $R$ of the star are determined when the above boundary condition is met.
Finally, the boundary condition on $e^{\nu}$ at $R$ is given by
\begin{equation}
    e^{\nu} = 1 - \frac{2 M}{R}.
\end{equation}

\subsection{Perturbation equations}

Let us next review the perturbation equations.
At $\mathcal{O}\left(\epsilon^1\right)$, the equation of motion reads
\begin{align}
    \dv[2]{\omega_1}{r} &= 4\frac{\pi r^2 \left( \rho + p_r \right) e^\lambda - 1}{r}\dv{\omega_1}{r} \nonumber\\
    &+ 16 \pi \left( \rho + p_r - \sigma_0 \right) e^\lambda \omega_1,
    \label{eq: 1st order perturbation}
\end{align}
and at $\mathcal{O}\left(\epsilon^2\right)$, the equation of motion reads
\begin{widetext}
\begin{align}
    \sigma_2^{(2)} &= \left( \rho + p_r - \sigma_0 \right) h_2 - \frac{\sigma_0}{r - 2 m} m_2 + \left[ \frac{\left( \rho + p_r \right) \left( 4 \pi p_r r^3 + m \right)}{r \left( r - 2 m \right)} + \dv{\sigma_0}{r} + 2 \frac{\sigma_0}{r} \right] \xi_2 + \frac{r^2}{3} \left( \rho + p_r - \sigma_0 \right) e^{-\nu} \omega_1^2, \label{eq: 2nd order perturbation sigma2}\\
    m_2 &= - r e^{-\lambda} h_2 + \frac{r^4}{6} e^{-\left( \nu + \lambda \right)} \left[ r e^{-\lambda} \left( \dv{\omega_1}{r} \right)^2 + 16 \pi r \omega_1^2 \left( \rho + p_r - \sigma_0 \right) \right], \\
    \dv{h_2}{r} &= - \frac{4 \pi p_r r^3 - m + r}{r} e^\lambda \dv{k_2}{r} + \frac{3}{r} e^\lambda h_2 + \frac{2}{r} e^\lambda k_2 + \frac{1 + 8\pi p_r r^2}{r^2} e^{2 \lambda} m_2 + \frac{r^3}{12} e^{-\nu}{\left( \dv{\omega_1}{r} \right)}^2 \nonumber\\
    &+ 4 \pi \left( \rho + p_r \right) \frac{4 \pi p_r r^3 + m}{r} e^{2 \lambda} \xi_2 + 8\pi e^\lambda \sigma_0 \xi_2, \label{eq: 2nd order perturbation h2}\\
    \dv{k_2}{r} &= - \dv{h_2}{r} - \frac{4 \pi p_r r^3 + 3 m - r}{r^2} e^\lambda h_2 + \frac{4 \pi p_r r^3 - m + r}{r^3} e^{2\lambda} m_2, \label{eq: 2nd order perturbation k2}\\
    \dv{\xi_2}{r} &= \frac{r - 2 m}{6 r \left[ \left( \rho + p_r \right) \left( 4 \pi p_r r^3 + m \right) + 2 \left( r - 2 m \right) \sigma_0 \right]} \biggl\{ - 6 r^2 \left( \rho + p_r \right) \dv{h_2}{r} -12 \sigma_0 r^2 \dv{k_2}{r} \nonumber\\
    &- 3 \left[ r^2 \left( \rho + p_r \right) \dv[2]{\nu}{r} - 4 \sigma_0 \right] \xi_2 - 12 r \sigma_2^{(2)} + 2 r^3 \left( \rho + p_r - \sigma_0 \right) e^{-\nu} \omega_1 \left[ \left( r \dv{\nu}{r} - 2 \right) \omega_1 - 2 r \dv{\omega_1}{r} \right] \biggl\}.
    \label{eq: 2nd order perturbation xi2}
\end{align}
\end{widetext}

The boundary conditions near $r=0$ are given by
\begin{align}
    \omega_1 &= \omega_{1c} + \frac{8}{5}\pi (p_c+ \rho_c) \omega_{1c} r^2+\mathcal{O}(r^3), \\
    h_2 &=   C_1 r^2 + \mathcal{O}(r^3), \\
    k_2 &= - C_1 r^2 + \mathcal{O}(r^3), \\
    m_2 &= - C_1 r^3 + \mathcal{O}(r^4), \\
    \xi_2 &= -\frac{\left(3 C_1 \,e^{\nu_c}+\omega_{1c}^{2}\right) \left(p_c +\rho_c \right)}{2 \left[2\pi (3 p_c+ \rho_c) (p_c + \rho_c)+3 \sigma_{02} \right] e^{\nu_c}} r \nonumber\\
    &+ \mathcal{O}(r^2), \\
    \sigma_{2}^{(2)} &= -\frac{\left(3 C_1 e^{\nu_c}+\omega_{1c}^{2}\right) \left(p_c + \rho_c \right) \sigma_{02}}{\left[2\pi (3 p_c+ \rho_c) (p_c + \rho_c)+3 \sigma_{02} \right] e^{\nu_c}} r^2 \nonumber\\
    &+ \mathcal{O}(r^3),
\end{align}
where $\sigma_{02}$ is a Taylor coefficient given by
\begin{equation}
    \sigma_0 = \sigma_{02} r^2 + \mathcal{O}(r^3).
\end{equation}

Together with the background configuration of the elastic star, we integrate Eq.~\eqref{eq: 1st order perturbation} and Eqs.~\eqref{eq: 2nd order perturbation h2}--\eqref{eq: 2nd order perturbation xi2} to obtain the interior solution of $\omega_1$, $h_2$, $k_2$, and $\xi_2$, respectively. 
The constants $\omega_{1c}$ and $C_1$ are determined by matching the interior and exterior solutions on the stellar surface at $R$.
Subsequently, the moment of inertia $I$ and spin-induced quadrupole moment $Q$ of a slowly rotating, isolated star can be extracted from the exterior solutions $\omega_1^{\mathrm{ext}}$ and $h_2^{\mathrm{ext}}$, respectively,  
\begin{align}
    \omega_1^{\mathrm{ext}} &= \Omega\left(1-\frac{2 I}{r^{3}}\right), \\
    h_2^{\mathrm{ext}} &= -\frac{Q}{r^{3}} + \mathcal{O} \left( \frac{1}{r^4} \right).
\end{align}
Meanwhile, the tidal deformability $\lambda_2$ of a nonrotating, tidally deformed star can be calculated by solving Eqs.~\eqref{eq: 2nd order perturbation sigma2}--\eqref{eq: 2nd order perturbation xi2} with $\omega_1 =0$. 
$\lambda_2$ is related to the tidal Love number $k_2^\mathrm{tid}$ as
\begin{equation}
    \lambda_2 = \frac{2}{3} k_2^{\mathrm{tid}} R^5.
\end{equation}
$k_2^\mathrm{tid}$ can be calculated from
\begin{equation}
\begin{aligned}
    k_2^{\mathrm{tid}} &= \frac{8}{5} C^{5}(1 - 2 C)^{2}[2 + 2 C (y-1)-y] \\
    &\times \bigl\{ 2 C[6-3 y+3 C(5 y-8)] \\
    &+ 4 C^{3} \left[ 13-11 y+C(3 y-2)+2 C^{2}(1+y) \right] \\
    &+ 3(1-2 C)^{2}[2-y+2 C(y-1)] \ln (1-2 C) \bigl\}^{-1},
\end{aligned}
\end{equation}
where
\begin{equation}
    y = \frac{r}{h_2^{\mathrm{ext}}} \dv{h_2^{\mathrm{ext}}}{r} \bigg|_{r=R}.
\end{equation}
We refer the readers to Refs.~\cite{Hartle:1967he,Hartle:1968si,Yagi:2013awa,Yagi:2015hda} for more detail on matching the interior and exterior solutions of $\omega_1$, $h_2$, $k_2$, and $\xi_2$. 
The above set of equations reduces to the background and perturbation equations for fluid stars with isotropic pressure~\cite{Yagi:2013awa}, i.e., $p_r = p_t = p$, when $\sigma_0$ vanishes. 

In this paper, the numerical integration of Eqs.~\eqref{eq: TOV(reformulated) m}--\eqref{eq: TOV(reformulated) z}, \eqref{eq: 1st order perturbation}, and \eqref{eq: 2nd order perturbation h2}--\eqref{eq: 2nd order perturbation xi2} is performed using the explicit fourth-order Runge-Kutta method. 
We verify that the fluid limit results are recovered when $\tilde{\mu}$ is taken to be an infinitesimally small value. 
Also, we compare our numerical results with Refs.~\cite{Yagi:2013awa,Yagi:2015hda,Dong:2024lte} to validate our code implementation.
We obtain consistent results when the same physical models are used.



\subsection{Junction condition on $\xi_2$}


Similar to $\tilde{p}$, the quantity $\xi$ is also discontinuous at the core-envelope interface due to the jump in $\sigma_0$ [see Eqs.~\eqref{eq: 2nd order perturbation sigma2} and \eqref{eq: 2nd order perturbation xi2}]. 
Although its value inside the fluid envelope can be found directly by setting $\sigma_0 = \sigma_2^{(2)} = 0$ in Eq.~\eqref{eq: 2nd order perturbation sigma2}, we here derive the junction condition that determines its jump for completeness.

First, we can write discontinuous $\sigma_0$ as
\begin{align}
    \sigma_0 = \Theta(r_t - r)\sigma_0^{-} + \Theta(r - r_t)\sigma_0^{+},
\end{align}
where $\Theta(x)$ is the Heaviside step function with the half-maximum convention [i.e., $\Theta(0) = 1/2$], the subscript $t$ represents that the quantity is evaluated at the transition point, and the superscript $-$ $(+)$ represents the core (envelope) side value of a quantity that is discontinuous at $r_t$. 
In our model, $\sigma_0^{+} = 0$. 
However, we keep it unspecified here to make the final expression generic. 
The derivative in Eq.~\eqref{eq: 2nd order perturbation sigma2} is thus given by
\begin{align}
    \frac{d\sigma_0}{dr} &= \Theta(r_t - r)\frac{d\sigma_0^{-}}{dr} + \Theta(r-r_t)\frac{d\sigma_0^{+}}{dr} \nonumber\\
    &+ \left(\sigma_{0,t}^{+}-\sigma_{0,t}^{-}\right) \delta(r-r_t).\label{eq: dsigmadr interface}
\end{align}
Applying Eq.~\eqref{eq: dsigmadr interface} to Eq.~\eqref{eq: 2nd order perturbation xi2} and integrating, we obtain 
\begin{align}
    \frac{\xi_{2,t}^{+} - \xi_{2,t}^{-}}{\xi_{2,t}^{+} + \xi_{2,t}^{-}} = \frac{2}{r_t} \left(\sigma_{0,t}^{+} - \sigma_{0,t}^{-}\right) \left(\frac{dp_{r}^{+}}{dr} + \frac{dp_{r}^{-}}{dr}\right)^{-1}.
    \label{eq: xi2 junction condition}
\end{align}
In practical calculations, we compute $\xi_{2,t}^{+}$ by setting $\sigma_0 = \sigma_2^{(2)} = 0$ in Eq.~\eqref{eq: 2nd order perturbation sigma2}, and the numerical results are consistent with that obtained using Eq.~\eqref{eq: xi2 junction condition}.

\bibliography{apssamp}
\end{document}